\documentclass[12pt]{article}

\usepackage{graphicx} 
\begin{document}

\noindent{\bf Perspective}

\bigskip

\begin{center}

{\large\bf Physics-Informed and Data-Driven Discovery of Governing Equations 
for Complex Phenomena in Heterogeneous Media}

\bigskip
Muhammad Sahimi

{\it Mork Family Department of Chemical Engineering and Materials Science,
University of Southern California, Los Angeles, California 90089-1211, USA}

\end{center}

\bigskip

Rapid evolution of sensor technology, advances in instrumentation, and progress
in devising data-acquisition softwares/hardwares are providing vast amounts of 
data for various complex phenomena, ranging from those in atomospheric 
environment, to large-scale porous formations, and biological systems. The 
tremendous increase in the speed of scientific computing has also made it 
possible to emulate diverse high-dimensional, multiscale and multiphysics 
phenomena that contain elements of stochasticity, and to generate large volumes
of numerical data for them in heterogeneous systems. The difficulty is, 
however, that often the governing equations for such phenomena are not known. 
A prime example is flow, transport, and deformation processes in
macroscopically-heterogeneous materials and geomedia. In other cases, the 
governing equations are only partially known, in the sense that they either 
contain various coefficients that must be evaluated based on data, or that they
require constitutive relations, such as the relationship between the stress 
tensor and the velocity gradients for non-Newtonian fluids in the momentum
conservation equation, in order for them to be useful to the modeling. Several 
classes of approaches are emerging to address such problems that are based on 
machine learning, symbolic regression, the Mori-Zwanzig projection operator
formulation, sparse identification of nonlinear dynamics, data assimilation,
and stochastic optimization and analysis, or a combination of two or more of 
such approaches. This Perspective describes the latest developments in this 
highly important area, and discusses possible future directions.

\newpage

\begin{center}
{\bf I. INTRODUCTION}
\end{center}

A wide variety of systems of scientific, industrial, and societal importance 
represent heterogeneous, and multiphase and multiscale media. Examples vary 
anywhere from large-scale porous formations, to composite materials, biological
systems, and the Earth's atmosphere. Many complex phenomena also occur in such 
systems, including fluid flow, transport, reaction, and deformation. Given the 
extreme importance of such systems to human and societal progress, the goal for
decades has been developing models that describe not only the multiscale and 
multiphase systems themselves, but also the phenomena that occur there.

Consider, as an example, the problem of air pollution in large urban areas.
Chemical oxidants, especially ozone, are major products of photochemical 
oxidation (reactions that are influenced by Sun) of primary pollutants emitted 
from various sources in the tropospheric layer [1]. Although the presence of 
ozone in the stratospheric layer is responsible for continuation of life on 
Earth, its presence in the troposphere is dangerous to humans' health and 
damaging to national and international economies [2]. Effective control of 
the pollutants requires accurate and comprehensive knowledge of the rates of 
emission and transport of the reactants that are present in the atmosphere, and
the chemical reactions that they participate in. In particular, since in the 
presence of nitrogen oxides, NO$_{\rm x}$, ozone production increases very 
significantly [1], which is the main cause of the formation of photochemical 
smog, detailed information on its concentration is needed for its control. 

Such data are continuously collected by a large number of sensors in large 
urban areas around the world, and have been becoming available. To analyze and 
understand the huge volume of the data that are being continuously collected, 
modeling of such phenomena has been pursued for decades. Large urban areas are,
however, highly complex media. Consider, for example, the Greater Tehran area, 
Iran's Capital, which begins on the tall Alborz mountains in the north, and 
ends in the desert in the south, or the Greater Los Angeles area that is 
sandwiched between San Bernardino, San Gabriel, and San Fernando mountains and 
the Pacific Ocean. Clearly, the terrains and topography of such large uraban 
areas are highly rough and complex. Any modeling of atmospheric pollution over 
the two areas must take into acount not only the effect of the large rough 
terrains - about 1300 and 87,000 km$^2$ for, respectively, the Greater Tehran 
and Los Angeles areas - and their rough topography, but also the dynamic 
changes that occur there continously on hourly, daily, monthly, and seaonal 
bases, as the two areas represent multiscale systems, not only in space, but 
also in time, which span at least 10-15 orders of magnitude. As a result, 
numerical simulation of such complex phenomena, even if the governing 
equations are known, is extremely difficult, as it involves turbulent flow, 
reactions with highly nonlinear kinetics, a huge number of reactants - 
typically five dozens or more - and the reaction products - over one hundred - 
the presence or absence of a wind velocity field, boundary conditions that vary
dynamically, and many other complicating factors [1,3].

The availability of vast amount of data is not limited to the problem of 
atmospheric pollution. It is estimated that, over the next decade, hundreds of 
billions of sensors that include airborne, seaborne and satellite remote 
sensing will be collecting vast amounts of data for many phenomena, such as
vegetation and plantation, and the characteristics of draught-stricken areas, 
an increasingly important problem worldwide. The same is also true of large 
geomedia and such complex problems as seismology, fracture propagation, and 
earthquakes. Analysis of such data, particularly those for which the 
signal-to-noise ratio is low, i.e., noisy data, understanding the subtle 
insights that they may provide, and incorporating them into accurate physical 
models is a Herculean task, requiring a paradigm shift.

Such a paradigm shift has slowly begun to emerge, with two classes of 
approaches are currently being developed. One class exploits deep-learning 
(DP), and more generally machine-learning (ML), algorithms in order to address 
the problem. The approach has been motivated by the fact that in many cases, 
the ML algorithms [4,5] are capable of extracting important features from vast 
amounts of data that are characterized by spatial and temporal coverage; see 
for example, Reichstein {\it et al.} [6]. In some cases, the governing 
equations for the complex phenomena for which the data have been collected may 
be known, which are then incorporated into a ML approach in order to develop 
predictive tools for studying the phenomena over spatial and time scales well
beyond those over which the existing data have been collected. Two 
representative examples of such approaches are the work of Kamrava {\it et al.}
[7] for modeling fluid flow in porous materials, and that of Alber {\it et al.}
[8] for modeling of biophysical and biomedical systems. In other cases, the
governing equations may be known, but they contain transpot and other types of 
coefficients that depend on the morphology of the systems in which the complex
phenomena occur, or require constitutive relationships, without which the
governing equations would not be very useful, unless one resorts to pure
empiricism. As a results, the constitutive relationships and/or the 
coefficients that the governing equations contain must be {\it discovered}.

The second class of approaches is intended for the systems in which the 
governing equations for physical phenomena occuring in them and, hence, for 
the associated data, are not known. Thus, one attempts to discover the 
equations using the large amount of data currently available. The lack of 
governing equations is particularly true for those phenomena that involve 
multiscale heterogeneity in the form of some sort of stochasticity. The 
discovery of such equations has dominated physical sciences and engineering for
the past several decades, as they provide predictions for system behavior. 

The classical approach has been based on the fundamental conservation laws, 
namely, the equations that describe mass, momentum and energy conservation. If 
a system is heterogeneous, the microscale conservation laws are averaged over 
an ensemble of its possible realizations in order to derive the macroscale 
equations. This is, however, valid only if there is a well-defined 
representative elementary volume (REV) or scale, i.e., the volume or length
scale over which the heterogeneous system can be considered as macroscopically
homogeneous, so that it is stationary over length scales larger than 
the REV. 

But, what if the REV does not exist, or is larger than the size of the system,
in which case the system is non-stationary, i.e., the probability distribution 
functions (PDFs) of its probability vary spatially from region to region? 
Examination of many important systems indicates that non-stationarity is more 
like the rule, rather than the exception. A good example is natural porous 
media at large (regional) length scales. It is known [9] that the physical 
properties of such media, such as their permeability and elastic constant 
approximately follow non-stationary stochastic functions [10]. Thus, the 
question is, what are the governing equations for a flow, transport, and 
deformation processes in such media?

Oother obvious examples are biological, and nano- and neuroscience systems for 
which first-principle calculations are currently very difficult, if not 
impossible, to carry out, whereas data for them are becoming abundant and, in 
many cases, with exceptional quality. In addition, the tremendous increase in 
the computational power is making it possible to emulate the behavior of 
diverse and complex systems that are high-dimensional, multiscale, and 
stochastic. The question, then, is, how can we discover the governing equations
that not only honor and better explain the data, but also provide predictions 
for the future, or over much larger length and time scales? It should be clear
that the ability to discover the governing equations based directly on the data
is of paramount importance in many modern scientific and engineering problems. 

This Perspective describes the emerging field of physics-informed and 
data-driven (PIDD) modeling of multiscale, multiphysics systems and phenomena 
and, in particular, the approaches for discovering the governing equations for
given sets of data that represent the characteristics of complex phenomenon in 
heterogeneous media. We describe the emerging approaches, discuss their
strengths and shortcoming, and point out possible future directions in this
rapidly developing and highly significant research area.

\begin{center}
{\bf II. THREE TYPES OF SYSTEMS}
\end{center}

In general, the success of any PIDD approach for predicting the macroscopic
properties of complex phenomena that occur in multiscale heterogeneous media 
depends on the amount of available data, on the one hand, and the structure and
complexity of the system itself, on the other hand. Thus, let us divide the
systems of interests into three categories:

(i) Systems for which the governing equations for the physical phenomena of 
interest are known, but the available data are limited. For example, Darcy's 
law together with the Stokes' equation describe slow flow of Newtonian fluids 
in microscopically disordered, but macroscopically homogeneous porous media, 
while the convective-diffusion equation describes transport of a solute and 
mass transfer in the same media [9]. The three equations contain flow and
transport coefficients - the permeability and dispersion coefficient - which 
characterize flow and transport processes and, in principle, depend on the 
disordered morphology of the pore space. They must either be predicted, or 
computed, assuming a reasonable model of the pore space, or measured by 
careful experiments. In this case, the goal is to develop a PIDD approach in 
order to correlate the permeability and the dispersion coefficients with the
morphology of the pore space (see below).

(ii) In the extreme opposite to (i) are systems for which large amounts of data
are available, but the governing equations for the physical phenomena of 
interest at the macroscale are not known. Thus, the goal is developing a PIDD 
algorithm for understanding such systems and the data, as well as discovering 
the governing equations for the phenomena of interest.

(iii) In between (i) and (ii) are systems for which some reasonable amounts of 
data - not too large or too small - are available, and the physics of the 
phenomena of interest is also partially known. For example, any fluid flow is 
governed by the equations that describe mass and momentum conservation 
equations in terms of the stress tensor, but if the fluid is non-Newtonian, the
constitutive relationship that relates the stress tensor to the velocity field 
may not be known. Many systems of current interest belong to this category of
systems, but the number of systems that belong to class (ii) of systems is not 
only large, but is also increasing rapidly, due to the rapid advances in data 
gathering and observations. 

\begin{center}
{\bf III. DATA ASSIMILATION}
\end{center}

Let us first describe data assimilation, which is a well-established concept 
that has been utlized in the investigations of the atmospheric and geoological 
sciences to make concrete predictions for weather, oceans, climate, and
ecosystems, as well as for properties of geomedia. Since data assimilation 
techniques improve forecasting, or help developing a more accurate model that 
provides us with a deeper understanding of such complex systems, they play an 
important role in studies of climate change, pollution of environment, and 
oceans, as well as geological systems.

Data assimilation combines observational data with the dynamical principles, or
the equations or models that govern a system of interest in order to obtain an 
estimate of its state that is more accurate than what is obtained by using only
the data or the physical model alone. Thus, in essence, data assimilation is
suitable for the first type of systems described in Sec. III., i.e., those for 
which some reasonable amounts of data are available, and the physics of the 
phenomena of interest is at least partially known. Both the data and the models
have errors, however. As discussed by Zhang and Moore [11], the errors of the 
data include are of random, systematic, and representativeness types. Models 
also produce errors because, often, they are simplified, or are incomplete to 
begin with, in order to make the computations affordable, which in turn 
generates error.

We do not intend to review in detail data assimilation methods, as they are 
well known. Therefore, we only mention and describe them briefly, since later
in this Perspective we show how data assimilation methods can be combined with
a machine-learning algorithm in order to not only improve forecasting, but
also reduced the computational burden significantly. 

There are at least four approaches to data assimilation, which are the Cressman
and optimal interpolation methods, three- or four-dimensional variational 
analysis, and the Kalman filter. They all represent least-squares methods, with
the final estimate selected in such a way as to minimize its uncertainty. In 
all four approaches, the set of data representing a system's state is denoted 
by {\bf x}. The actual or true state ${\bf x}_t$ is different from the best 
possible representation ${\bf x}_b$ produced by physical models and referred to
as the background state. To analyze the system and the data, an observation 
vector {\bf y} is compared with the state vector. 

In Cressman method [12], which belongs to a class of methods called objective 
analysis, one assumes that the model state is univariate and is represented by 
values of the variable at discrete grid points. Suppose that a previous 
estimate of the model state is an $n$-dimensional vector, ${\bf  x}_b =[x_b(1),
\cdots, x_b(n)]^{\rm T}$, while the observed vector is an $n$-dimensional
vector, ${\bf y}=[y(1),\cdots,y(n)]^{\rm T}$. The Cressman method gives an 
updated model, ${\bf x})a=[x_a(1),\cdots,x_a(n)]^{\rm T}$, by the following 
equation
\begin{equation}
{\bf x}_a(j)={\bf x}_b(j)+\frac{\displaystyle \sum_{i=1}^n\omega_{ij}[{\bf y}
(i)-{\bf x}_b(i)]}{\displaystyle \sum_{i=1}^n\omega_{ij}}\;,
\end{equation}
with, $\omega_{ij}={\rm max}[0,(R^2-d_{ij}^2)/(R^2+d_{ij}^2)]$, and $d_{ij}=
|i-j|$. Note that, $\omega_{ij}=1$, if $i=j$, and $\omega_{ij}=0$, if $d_{ij}>
R$. $R$, which is a control parameter defined by the user, is referred to as
the influence parameter.

In the optimal interpolation method one combines the observation vector {\bf y}
with $p$ entries with the background vector ${\bf x}_b$ wth $n$ entries, with
$n\geq p$. Because there are usually fewer observations than variables in the
background model, the only correct way of making the comparison is to use an 
observation operator $h$ from model $n$-dimensional state space to 
$p$-dimensional observation space, which is a $p\times n$ matrix {\bf H} such
that $h({\bf x}_b)=(h_1,\cdots,h_p)^{\rm T}={\bf H}{\bf x}_b$, with $h_i=
\sum_{j=1}^n H_{ij}x_b(j)$.

Suppose that {\bf B} of size $n\times n$ and {\bf R} with a size $p\times p$ 
are, respectively, the covariance matrices of the background error ${\bf 
x}_b-{\bf x}_t$, and observation error ${\bf y}-h({\bf x}_b)$. The two errors
are assumed to be uncorrelated. The $n$-dimensional analysis, or updated, 
vector ${\bf x}_a$ is defined by, ${\bf x}_a={\bf x}_b+{\bf w}[{\bf y}-h({\bf
x}_b)]$, where {\bf w} is an $n\times p$ matrix that is selected such that
the variance of ${\bf x}_a-{\bf x}_t$ is minimized. It can be shown that,
${\bf w}={\bf BH}^{\rm T}({\bf HBH}^{\rm T}+{\bf R})^{-1}$.

In the three-dimensional variational analysis, a cost function $\sigma^2
({\bf x})$ is defined by
\begin{equation}
\sigma^2({\bf x})=({\bf x}-{\bf x}_b)^{\rm T}{\bf B}^{-1}({\bf x}-{\bf x}_b)+
[{\bf y}-h({\bf x})]^{\rm T}{\bf R}^{-1}[{\bf y}-h({\bf x})]\equiv \sigma_b^2
({\bf x})+\sigma_o^2({\bf x})\;,
\end{equation}
with $\sigma^2_b$ and $\sigma^2_o$ being the background and observation cost
functions. It has been proven that if we write, ${\bf x}={\bf x}_a={\bf x}_b+
{\bf w}[{\bf y}-h({\bf x})_b]$, then the cost function attains its global
minimum.

Generalization of the method to four-dimensional variational assimilation is 
straightforward. The observations are distributed among $(N+1)$ times in the 
interval of interest. The cost function is defined as
\begin{equation}
\sigma^2({\bf x})=({\bf x}-{\bf x}_b)^{\rm T}{\bf B}^{-1}({\bf x}-{\bf x}_b)+
\sum_{i=0}^N[{\bf y}_i-H_i({\bf x}_i)]^{\rm T}R_i^{-1}[{\bf y}_i-H_i({\bf 
x}_i)]\;,
\end{equation}
and, therefore, the data assimilation problem with globally minimum variance is
reduced to computing the analysis vector ${\bf x}_a$ such that $\sigma^2({\bf 
x})$ attains its minimal at x ${\bf x}={\bf x}_a$.

The Kalman filter [13], also known as linear quadratic estimation, has been 
used to continuously update the parameters of models of dynamical systems for 
assimilating data. The filter is optimal only under the assumption that the 
system is linear and the measurement and process noise follow Gaussian 
distributions. The algorithm, a recursive one, consists of two steps. In the 
prediction step, the filter generates estimates of the current state variables 
{\bf x}, together with their uncertainties. After the data for the next 
measurement, which may have some error, become available, step two commences 
in which the estimates are updated using a weighted average, with more weight 
given to estimates with greater certainty (smaller errors). The algorithm 
operates in real time, using only the present input measurements, and the state
calculated previously and its uncertainty matrix. The algorithm fails, however,
for highly nonlinear systems, which motivated the develpment of the extended 
Kalman filter by which the nonlinear characteristics of the system's dynamics 
are approximated by a version of the system that is linearized around the last 
state estimate. The extended version has been popular due to its ability for 
handling nonlinear systems and non-Gaussian noise. 

Evensen [14] identified a closure problem associated with the extended Kalman 
filter in the evolution equation for the error covariance. The problem in this
context is having more unknowns than equations. The linearization used in the 
extended filter discards higher-order moments in the equation that governs the 
evolution of the error covariance. But, because this kind of closure technique 
produces an unbounded error growth, the ensemble Kalman filter was introduced 
to alleviate the closure problem, which is a Monte Carlo method in which the 
model states and uncertainty are represented by an ensemble of realizations of 
the system [15]. 

The ensemble Kalman filter is conceptually simple and requires relatively low 
computation, which is why it has gained increasing attention in history 
matching problems and continuous updating of models, as new data become 
available. Since, instead of computing the state covariance using a recursive 
method, the method estimates the covariance matrix from a number of 
realizations, its computational cost is low. The ensemble Kalman filter has 
been shown to be very efficient and robust for real-time updating in various 
fields, such as weather forecasting [16], oceanography, and meteorology [17]. 
It was also used in the development of dynamic models of large-scale porous 
media [18] and optimizing gas production from large landfills [19], in both of 
which dynamic data become available over a period of time. The reader is 
referred to Ref. [19] for complete details of the method and how it is 
implemented.

\begin{center}
{\bf IV. PHYSICS-INFORMED MACHINE-LEARNING APPROACHES}
\end{center}

Machine-learning algorithms, and in particular neural networks, have been
used for decades to predict properrties of various types of systems [20], after
training the networks with some data. The problem that many machine-learning 
algorithms suffer from is that, they lack a rigorous, physics-based foundation 
and rely on correlations and regression. Thus, although they can fit very 
accurately a given set of data to some functional forms, they do not often have
predictive power, particularly when they are tasked with making predictions for
systems for which no data were ``shown'' to them, i.e., none or very little 
data for the properties to be predicted were used in training the NNs. 

This motivated the development of physics-informed machine-learning (PIML) 
algorithms, which are those in which, in addition to providing a significant 
amount of data for training the network, some physical constraints are also 
imposed on the algorithms. For example, if macroscopic properties of 
heterogeneous materials, such as their effective elastic moduli, are to be 
predicted by a neural network, then, in addition to the data that are used for 
training it, one can also impose the constraint that the predictions must 
satisfy rigrous upper and lower bounds derived for the moduli [21,22]. Or, if 
one is to use a machine-learning algorithm to predict fluid flow and transport 
of a Newtonian fluid in a porous medium, one can impose the constraint that the
training must include the Navier-Stokes equation, or the Stokes' equation if 
fluid flow is slow, and the convective-diffusion equation if one wishes to 
predict the concentration profile of a solute in the same flow field. Any other
constraint that is directly linked with the physics of the phenomenon may also 
be imposed.

The available data can then be incorporated into a machine-learning algorithm 
to link the structure of the system to the coefficients that appear in the 
equations that are known to govern the phenomena, and/or to discover the 
constitutive relations that are required for solving the governing equations. 
For example, a deeo-learning algorithm was used to link the morphology of 
porous media to their permeability [23] and the dispersion coefficient [24] in 
slow flow through the same pore space, as well as the diffusivity [25] and 
other propertties [26,27]. In addition, the same type of approaches have been 
used for developing a mapping between the conductivity field and the 
longitudinal macrodispersion coefficient in a 2D Gaussian field in porous 
media [28].

In general, three distinct approaches are being developed that contribute to 
the accuracy and acceleration of the training of a PIML algorithm that are as 
follows [4,5,7,23,29,30].

\begin{center}
{\bf A. Multi-Task Learning}
\end{center}

In this approach, the cost function, which is minimized globally in order to
develop the optimal machine-learning algorithm, and the neural network 
structure include the aforementioned constraints. In other words, it is not 
enough for the traditional cost function of the neural networks - the sum of 
the squares of the differences between the predictiona and the data - to be 
globally minimum, but rather the cost function is penalized by imposing the 
constraints on it. Thus, the approach is a multi-task learning process, because
not only the PIML algorithm is trained by the data, but the training also 
includes some physics-based constraints, such as a governing equation, upper 
and/or lower bounds to the properties of interest, and other rigorous 
information and insights, so that the predictions will also be based on, and 
satisfy, the constraints. The imposition of the constraints represents biases 
in the training process, as the constraints force the algorithm to be trained 
in a specific direction. We present two concrete examples to illustrate the 
method. 

{\it Example 1: predicting fluid flow in a thin, two-dimensional (2D) polymeric
porous membrane.} A high-resolution 3D image of the membrane of size $500\times
500\times 1000$ voxels was used [7], whose porosity, thickness, permeability, 
and mean pore size were known. Seven hundred 2D slices with a size $175\times 
175$ pixels were extracted from the 3D image, and fluid flow in the slices was 
simulated by solving the Navier-Stokes equations, with part of the results used
in the training the algorithm. 

A physics-informed recurrent encoder–decoder (PIRED) network was then 
developed. The network, a supervised one, consisted of encoder and decoder, 
known as the U-Net and residual U-Net (RU-Net), whose architecture is shown in 
Fig. 1. The encoder had four blocks, with each block containing the standard 
convolutional and activation layers, as well as pooling and batch normalization
layers. The pooling layer compressed the input images to their most important 
features by eliminating the unnecessary ones, and stored them in the latent 
layer that consisted of the activation, convolutional, and batch normalization
layers. The bath normalization layer not only allowed the use of higher 
learning rates by reducing internal covariate shift, but also acted as a 
regularizer for reducing overfitting [31]. The mean $\langle x\rangle$ and 
variance Var[$x$] of batches of data $x$ were computed in the bath 
normalization layer, and a new normalized variable $y$ was defined by
\begin{equation}
y=\gamma\frac{x-\langle x\rangle}{\sqrt{{\rm Var}[x]+\epsilon}}+\beta\;.
\end{equation}
Here, $\gamma$ and $\beta$ are learnable parameter vectors that have the same
size as the input data, and $\epsilon$ is set at a typically small value, 
$10^{-5}$ in this case. During the training, the layer kept running estimates 
of its computed mean and variance, and utilized them for normalization during 
evaluation. The variance was calculated by the biased estimator. 

The decoder also had four blocks. Each block contained the convolutional, 
activation and batch normalization layers, as well as a transposed 
convolutional layer that is similar to a deconvolutional layer in that, if, for
example, the first encoder has a size $128\times 64\times 64$, i.e., 128 
features with a size $64\times 64$, then, one has a similar size in the 
decoder. The transposed convolutional layer utilized the features extracted by
the pooling layer to reconstruct the output, which were the pressure and fluid 
velocity fields, $P$ and {\bf v}, at various times. Because the latent layer of
the recurrent neural network consisted of residual blocks, i.e., layers that, 
instead of having only one connection, were connected to more distant previous 
layers, it improved the performance of the PIRED, and sped up significantly the
overall network's computations.

Assuming that the fluid is incompressibe and Newtonian, the mass conservation 
equation for a 2D medium is given by, $\mbox{\boldmath$\nabla$}\cdot{\bf v}=
\partial v_x/\partial x+\partial v_y/\partial y=0$, where both velocity 
components $v_x$ and $v_y$ and the spatial coordinates $x$ and $y$ are made 
dimensionless by a characteristic length $L$ and characteristic velocity $v_0$.
The (dimensionless) Navier-Stokes equation is given by 
\begin{equation}
\frac{D{\bf v}}{Dt}=\frac{\partial{\bf v}}{\partial t}+{\bf v}\cdot
\mbox{\boldmath$\nabla$}{\bf v}={\rm Re}^{-1}\left(-\mbox{\boldmath$\nabla$}P
+\nabla^2{\bf v}\right)\;,
\end{equation}
where ${\rm Re}=\rho v_0L/\mu$ is the Reynolds number, and $D/Dt=\partial/
\partial t+{\bf v}\cdot\mbox{\boldmath$\nabla$}$. Three residual functions, 
$\xi_1=\mbox{\boldmath$\nabla$}\cdot{\bf v}$, $\xi_2=Dv_x/Dt-{\rm Re}^{-1}\left
(-\partial P/\partial x+\nabla^2v_x\right)$, and $\xi_3=Dv_y/Dt-{\rm Re}^{-1}
\left(-\partial P/\partial y+\nabla^2v_y\right)$, were defined and incorporated
in the cost function $\sigma^2$, minimized by the PIRED network, instead of 
naively minimizing the squared differences between the data and predicted 
values of {\bf v} and $P$. To converge to the actual, numerically calculated 
values by solving the mass conservation and the Navier-Stokes equations, one 
must have, $\xi_i=0$ for $i=1-3$. Thus, the PIRED network learned that the 
mapping between the input and output must comply with the requirement that, 
$\xi_i=0$, which not only enriched its training, but also accelerated 
convergence to the actual values. The cost function $\sigma^2$ was, therefore, 
defined by
\begin{equation}
\sigma^2=\frac{1}{n}\left\{\sum_{i=1}^n\left[(P_i-\hat{P}_i)^2+(|v_i|-
|\hat{v}_i|)^2\right]\right\}+\sum_{i=1}^3\sum_{j=1}^n\xi_i(x_j,y_j,t_j)^2\;,
\end{equation}
where $n$ is the number of data points used in the training, and $P_i$ and 
$|v_i|$ are the actual pressure and magnitude of the fluid velocity at point 
$(x_i,y_i)$ at time $t_i$, with superscript $\hat{}$ denoting the predictions
by the PIRED network. The $P$ and {\bf v} fields were computed at four distinct
times. Note that the amount of the data needed for computing $P$ and {\bf v} 
was significantly smaller than what would be needed by the standard 
machine-learning methods.

The fluid was injected at one side and a fixed pressure was applied to the 
opposite side of the membrane. The other two boundaries were assumed to be 
impermeable. Solving the mass conservation and Navier-Stokes equations in each 
2D image took about 6 CPU minutes. The computations for training the PIRED 
network on an Nvidia Tesla V100 graphics processing unit (GPU) took about 2 GPU
hours. Then, the tests for accuracy took less than a second. Part of the 
results were used in the training, and the rest in testing and making 
comparison with the predictions of the PIRED network.

The reverse Kullback-Leibler divergence (relative entropy) [32] was used 
to minimize the cost function $\sigma^2$. If $p(x)$ is the true probability 
distribution of the input/output data, and $q(x)$ is an approximation to 
$p(x)$, the reverse Kullback-Leibler divergence from $q(x)$ to $p(x)$ is a 
measure of the difference between the two. The aim is, of course, to ensure 
that $q(x)$ represents $p(x)$ accurately enough that it minimizes the reverse 
Kullback-Leibler divergence $D_{\rm KL}(q\|p)$, defined by
\begin{equation}
D_{\rm KL}[q(x)\|p(x)]=\sum_{x\in S}q(x)\log\left[\frac{q(x)}{p(x)}\right]\;,
\end{equation}
where $S$ is the space in which $p(x)$ and $q(x)$ are defined. $D_{\rm KL}=0$, 
if $q(x)$ matches $p(x)$ perfectly and, in general, it may be rewritten as
\begin{equation}
D_{\rm KL}[q\|p]=E_{x\sim q}[-\log p(x)]-H[q(x)]\;,
\end{equation}
where $H[q(x)]=E_{x\sim q}[-\log q(x)]$ is the entropy of $q(x)$, with $E$
denoting the expected value operator and, thus, $E_{x\sim q}[-\log p(x)]$ being
the cross-entropy between $q$ and $p$. Optimization of $D_{\rm KL}$ with 
respect to $q$ is defined by
\begin{equation}
\arg\min D_{\rm KL}[q\|p]=\arg\min E_{x\sim q}[-\log p(x)]-H[q(x)]=\arg\min 
E_{x\sim q}[\log p(x)]+H[q(x)].
\end{equation}
Thus, according to Eq. (9), one samples data points from $q(x)$ and does so 
such that they have the maximum probability of belonging to $p(x)$. The entropy
term of Eq. (9) ``encourages'' $q(x)$ to be as broad as possible. The 
autoencoder tries to identify a distribution $q(x)$ that best approximates 
$p(x)$.

The trained PIRED network was used to reconstruct the velocity and pressure 
field in new (unused in training) 2D images using only a small number of 
images. Figures 2(a) and 2(b) present, respectively, the change in the cost 
function $\sigma^2$ for the training and testing datasets of the network. 
$\sigma^2$ decreases for both $P$ and {\bf v} during both the training and 
testing, indicating convergence toward the true solutions for both the pressure
and fluid velocity fields. 

An effective permeability $K$ was defined by, $K=\mu Lq/(A\Delta P)$, where 
$q$, $A$ and $\Delta P$ are, respectively, the steady-state volume flow rate, 
and the surface area perpendicular to the macroscopic pressure drop $\Delta P$.
$K$ was computed for 300 testing slices, and was predicted by the PIRED network
as well. The comparison is shown in Fig. 3(a). But a most stringent test of the
PIRED network is if one predicts the permeability (and other properties) of a 
completely different porous medium without using any data associated with it. 
Thus, the image of a Fontainebleau sandstone [33] with a porosity of 0.14 was 
used. Since the sandstone's morphology is completely different from the 
polymeric membrane's, a slightly larger number of 2D slices from the membrane 
(not the sandstone) was utilized to better train the PIRED network. Figure 3(b)
compares the effective permeabilities of one hundred 2D slices of the sandstone
with the predictions of the PIRED network. 

{\it Example 2: predicting arterial blood pressure in cardiovascular flows.}
Predictive modeling of cardiovascular flows and aspire is a valuable tool for 
monitoring, diagnosis and surgical planning, which can be utilized for large 
patient-specific topologies of systemic arterial networks, in order to obtain 
detailed predictions for, for example, wall shear stresses and pulse wave 
propagation. The models that were developed in the past relied heavily on 
pre-processing and calibration procedures that require intensive computations,
hence hampering their clinical applicability. Kissas {\it et al.} [34] 
developed a machine-learning approach, a physics-informed neural network (PINN)
for seamless synthesis of non-invasive in-vivo measurements and computational 
fluid dynamics.

Making a few assumptions, Kissas {\it et al.} [34] modeled pulse wave 
propagation in arterial networks by a reduced order (simplified) 1D model based
on the mass conservation and momentum equations,
\begin{eqnarray}
& & \frac{\partial A}{\partial t}+\frac{\partial(Av_x)}{\partial x}=0\;,\\
& & \frac{\partial v_x}{\partial t}+\alpha v_x\frac{\partial v_x}{\partial x}+
\left(\frac{v_x}{A}\right)\frac{\partial}{\partial x}[(\alpha-1)Av_x]+\frac{1}
{\rho}\frac{\partial P}{\partial x}-K_R\frac{v_x}{A}=0\;.
\end{eqnarray}
Here, $A(x,t)$, $v_x(x,t)$ and $P(x,t)$ denote, respectively, the 
cross-sectional area, blood's velocity, and pressure at time $t$, with $x$ 
being the direction of blood flow; $\alpha$ is a momentum flux correction 
factor; $\rho$ is the blood's density, and $K_R$ is a friction parameter that 
depends on the velocity profile (flow regime). However, since the artery is an 
elastic material that can be defomed, the constraint imposed by mass and 
momentum conservation is not sufficient for determining the pressure, since 
only the pressure gradient appears in the momentum equation. Assuming, However,
that the artery is a linearly elastic material, the constitutive law for 
displacement of its walls, given by
\begin{equation}
P=P_e+\beta(\sqrt{A}-\sqrt{A_0})\;,
\end{equation}
relates directly the arterial wall displacement to the absolute pressure in 
each cross section. Here, $\beta$ is a coefficient related to the Young's 
modulus and the Poisson's ratio of the artery; $A_0=A(x,0)$, and $P_e$ is the 
external pressure. Thus, as another constraint, the constitutive relation was 
coupled to the mass and momentum conservation laws, implying that the 
correlations between them can be exploited through the PINN in order to
determine the absolute pressure from velocity and cross-sectional area 
measurements. The system that Kassas {\it et al.} [34] modeled and studied, a 
$Y-$shaped bifurcation, is shown in Fig. 4. Three-dimensional geometries 
recovered from magnetic resonance imaging data and the corresponding 
center-lines (shown in Fig. 4) were extracted by using the vascular modeling 
toolkit library. The governing equations were then discretized and solved 
numerically by discontinuous Galerkin method.

Thus, similar to the first example described above, three residual functions, 
$\xi_i$ with $i=1,2,$ and 3, were defined as the left sides of Eqs. (10) - 
(12). Several factors contribute to the overall cost, or loss, function, 
$\sigma^2$, which should be minimized globally. They are, (a) the usual sum of 
the squared differences between the computed blood velocity and the arterial 
cross section and the corresponding data at every computational point $(x,t)$. 
Blood velocity data are typically obtained using Doppler ultrasound or 4D flow 
MRI, while the area data are gleaned from 2D Cine images recovered by 4D flow 
MRI. (b) The sum of the squared residual functions $\xi_i$, defined above, at a
sample of the collocation points used in the numerical simulation of mass 
conservation and momentum equations. (c) Contributions by the junctions at the 
bifurcation points. Consider Fig. 4. We refer to the channel on the left as 
artery 1, and the two on the right that bifurcate from it as numbers 2 and 3. 
Conservation of mass requires that, $A_1v_1-(A_2v_2+A_3v_3)=0$, where, for 
convecience, we deleted the subscript $x$ of the fluid velocities. Moreover, 
conservation of momentum implies that, $p_1+\frac{1}{2}\rho v_1^2-(p_2+\frac{1}
{2}\rho v_2^2)=0$, and $p_1+\frac{1}{2}\rho V_1^2-(p_3+\frac{1}{2}\rho v_3^2)=0
$. Thus, three additional residual functions, $\xi_i$ with $i=4,5,$ and 6 were 
defined by the left sides of the above equations, and the overall cost function
$\sigma^2$ was the sum of the three types of contributions.

In many problems of the type we discuss here, there maybe an additional 
complexity: The order of magnitude of fluid velocity, cross-sectional area and 
pressure are significantly different. For example, one has, $P\sim 10^6$ Pa, 
$A\sim 10^{-5}$ m$^2$, and $v_x\sim 10$ m/s. Such large differences give rise 
to a systematic numerical problem during the training of the PINN, since it 
affects severely the magnitude of the back-propagated gradients that adjust the
neural network parameters during training. To address this issue and similar to
Example 1 above, Kissas {\it et al.} made the governing equations dimensionless
by defining a characteristic length and a characteristic velocity, so that they
all take on values that are ${\cal O}(1)$. They then normalized the input to 
have zero mean and unit variance, since as Glorot and Bengio [35] demonstrated,
doing so mitigates the pathology of vanishing gradients in deep neural 
networks. The activation function that Kissas {\it et al.} utilized was a 
hyperbolic tangent function. 

Three neural networks, one for each artery in thye $Y-$shape system, were used.
Each of the networks had seven hidden layers with one-hundred neurons per 
layer, followed by hyperbolic tangent activation function. Two thousands 
collocation points were used in the discontinuous Galerkin method for solving 
the discretized equations. Other details of the approach and the model are 
given in the original reference.

Figure 5 presents the results in $Y-$shaped bifurcation. Figure 5(a) compares
the predicted velocity wave, computed by discontinuous Galerkin solution, with 
the predictions of the PINN with non-dimensionalization and without it, while 
Fig. 5(b) does the same for the pressure. They were computed at the middle 
point of artery 1. The agreement is excellent. The same type of approach was 
utilized by Zhu {\it et al.} [36] for surrogate modeling and quantifying 
uncertainty, and by Geneva and Zabaras [37] and Wu {\it et al.} [38] for 
modeling of nonlinear dynamical systems.

\begin{center}
{\bf B. Learning Aided by Physical Observations}
\end{center}

The training of any machine-learning algorithm can be improved by feeding it, 
as the input, observational data that convey the physics of the system under 
study. As mentioned in the Introduction, vast amounts of data are being 
collected for various complex phenomena. Thus, if such data, which provide 
insights into the phenomena are used as the input to training of a 
machine-learning algorithm, they will bias it toward satisfying the 
observational data, implying that the final machine-learning tool should be 
capable enough for providing accurate predictions for those aspects of the 
phenomenon for which no data were fed to the algorithm as the input; see, for 
example, Kashefi {\it et al.} [39] who developed a point-cloud deep-learning 
algorithm for predicting fluid flow in disordered media. A point cloud is a set
of data points that is typically sparse, irregular, orderless and continuous, 
encodes information in 3D structures, and is in per-point features that are 
invariant to scale, rigid transformation, and permutation. Due to such 
characteristics, feature extractions from a point cloud is difficult for many 
deep-learning models.

\begin{center}
{\bf C. Embedding Prior Knowledge and Inductive Biases}
\end{center}

One may design neural networks in which prior knowledge and inductive biases 
are embedded, in order to facilitate making predictions for the phenomena of 
interest. Convolutional neural networks, first proposed by LeCun {\it et al.} 
[40], are the best known examples of such approaches. They were originally 
designed such that the invariance along groups of symmetries and patterns found
in nature were honored. It has also been possible to design more general
convolutional neural networks that honor such symmetry groups as rotations and 
reflections, hence leading to the development of architectures that depend only
on the intrinsic geometry, which have been shown to be powerful tools for 
analyzing medical images [41] and climate pattern segmentation [42].

Kernel methods [43] in which optimization is carried out by minimizing the cost
function over a space of functions, rather than over a set of parameters as in 
the old neural network, is another approach that falls into the class of 
algorithms that improve the performance of the PIML approaches. They were 
motivated [43-45] by the physics of the systems under study. Moreover, many 
approaches that utilize neural networks have close asymptotic links to the 
kernel methods. For example, Wang {\it et al.} [46,47] showed that the training
dynamics of the PIML algorithms can be understood as a kernel regression method
in which the width of the network increases without bound. In fact, neural 
network-based methods may be rigorously interpreted as kernel methods in which 
the underlying {\it warping kernel} - a special type of kernels that were 
initially introduced [48] to model non-stationary spatial structures - is also 
learned from data. 

In many machine-learning processes, the training process must deal with data 
that are presented as graphs, which imply relations and correlations between 
the information that the graphs contain. Examples include learning molecular 
fingerprints, protein interface, classifying diseases, and reasoning on 
extracted structures, such as the dependency trees of sentences. Graph neural 
networks and their variants, such as graph convolutional networks, graph
attention networks, and graph recurrent networks, have been proposed for such 
problems, and have proven to be powerful tools for many deep-learning tasks. An
excellent review was given by Zhou {\it et al.} [49]; see also Refs. [7,29,30] 
for their applications.

It should be clear that one may combine any of the above three approaches in 
order to gain better performance of machine-learning algorithms. In addition, 
as the PIRED example described above demonstrated, when one deals with problems
involving fluid flow, transport, and reaction processes in heterogeneous media,
one may introduce dimensiolness groups, such as the Reynolds, Froude, and 
Prandtl numbers that not only contain information about and insights into the 
physics of the phenomena, but may also help one to upscale the results obtained
by the PIML algorithm to larger length and time scales. 

The field of PIML algorithms has been rapidly advancing. Many applications have
been developed, particularly for problems for which either advanced classical
numerical simulations pose extreme difficulty, or they are so ill-posed that 
render the classical methods useless. They include, in addition to those 
referenced above, PIML for 4D flow magnetic resonance imaging data [34], 
predicting turbulent transport on the edge of magnetic confinement fusion 
devices - a problem that has been studied for several decades [50] - and a 
fermionic neural network (dubbed FermiNet) for ab initio computation of the 
solution of many-electron Schr\"odinger equation [51,52] (see also Ref. [53]), 
which is a hybrid approach for informing the neural network about the physics 
of the problem. Since the wavefunctions must be parameterized, a special 
architecture was designed for FermiNN that followed the Fermi-Dirac statistics,
i.e., it was anti-symmetric under the exchange of input electron states and the
boundary conditions. As such, the parametrization was a physics-informed 
process. FermiNet was also trained by a physics-informed approach in that, the 
cost function was set as a variational form of the value of the energy 
expectation, with the gradient estimated by a Monte Carlo method. Several 
papers have explored application of the PIML to geoscience [4,7,23-26,28-30,53,
54], as well as to large-scale molecular dynamics simulations [55] in which a 
neural network is used to represent the potential energy surfaces, and 
pre-processing is used to preserve the translational, rotational and 
permutational symmetry of the molecular system. The algorithm can be improved 
by using deep potential molecular dynamics, DeePMD [56], which makes it 
possible to carry out molecular dynamics simulations with one hundred million 
atoms for more than one nanosecond long [57], as well as simulations whose 
accuracy was comparable with ab initio calculations with one million atoms 
[57,58].

\begin{center}
{\bf V. DATA-DRIVEN RECONSTRUCTION OF GOVERNING EQUATIONS}
\end{center}

As mentioned earlier, advances in technology and instrumentation have made it
possible to collect very large amounts of data for various phenomena in systems
that contain some type of heterogeneity, and the goal is to discover or
reconstruct the governing equations that describe such data. The approach that 
we describe in this section is suitable for the third type of systems discussed
in Sec. II, i.e., those in which extensive data are available for a given 
system, the governing equation is known, or is assumed so, but one must use a 
data-driven approach to reconstruct the equation by estimating its 
coefficients.

The approach has been developed for systems for which the data are in the form 
of nonstationary time series $X(t)$, or spatially-varying series $X({\bf x})$.
Characterizing such nonstationary time and spatial series has been a problem of
fundamental interest for a long time, as they are encountered in a wide variety
of problems, ranging from economic activity [59], to seismic time series [60], 
heartbeat dynamics [61,62], and large-scale porous media [9], and their
analysis has a long and rich tradition in the field of nonlinear dynamics 
[63-65]. Much of the effort has been focused on addressing the question of how 
to extract a deterministic dynamical system of equations by an accurate 
analysis of experimental data since, if successful, the resulting equations 
will yield all the important information about and insights into the system's 
dynamical properties. 

The standard approach has been to treat the fluctuations in the data as 
stochastic variables that have been superimposed {\it additively} on a 
trajectory or time series that the deterministic dynamical system generates.
The approach was originally motivated by the efforts for gaining deeper 
understanding of turbulent flows [66,67], and has been evolving ever since.
Although it has already found many applications [68], it is still under further
development (see below). More importantly, the approach has demonstrated the 
necessity of treating the fluctuations in the data as dynamical variables that 
interfere with the deterministic framework. 

In this approach, given a nonstationary series $X(t)$, one constructs a 
stationary process $y(t)$, which can be done by at least one of two methods. 
(i) The {\it algebraic increments}, $y(t)=X(t+1)-X(t)$, are constructed. The 
best-known example of such series is the fractional Brownian motion (FBM) [69] 
with a power spectrum, $S(\omega)\propto 1/\omega^{2H+1}$, where $H$ is the 
Hurst exponent. It is well-known that the FBM's increments, with $S(\omega)
\propto 1/\omega^{2H-1}$ and called fractional Gaussian noise [69], are 
stationary. Moreover, when $H=1/2$, the increments are uncorrelated, whereas 
for $H=-1/2$ $X(t)$ becomes random. (ii) Let $Z=\ln X(t)$. Then, one constructs
the {\it returns} $y(t)$ of $X(t)$ by, $y(t)=Z(t+1)-Z(t)=\ln[X(t+1)/X(t)]$, so 
that $y(t)$ is the {\it logarithmic increments} series. It is straightforward 
to show that both approaches yield stationary series by studying their various
moments over windows of different sizes in the series. One then analyzes $y(t)$
based on the application of Markov processes and derives a governing equation 
for the series based on a Langevin equation, the details of which are as 
follows.

One first checks whether $y(t)$ does follow a Markov chain [70,71]. If so, its 
Markov time scale $t_M$ - the minimum time interval over which $y(t)$ can be 
approximated by a Markov process - is estimated (see below). In general, to 
characterize the statistical properties of any series $y(t)$, one must evaluate
the joint probability distribution function $P_n(y_1,t_1;\cdots;y_n,t_n)$ for 
the number of the data points, $n$. If, however, $y(t)$ is a Markov process, 
the $n$-point joint probability distribution function $P_n$ is given by
\begin{displaymath}
P_n(y_1,t_1;\cdots;y_n,t_n)=\prod_{i=1}^{n-1}P(y_{i+1},t_{i+1}|y_i,t_i)
\end{displaymath}
where $P(y_{i+1},t_{i+1}|y_i,t_i)$ is the conditional probability. Moreover,
satisfying the Chapman-Kolmogorov equation [72],
\begin{equation}
P(y_2,t_2|y_1,t_1)=\int dy_3\;P(y_2,t_2|y_3,t_3)P(y_3,t_3|y_1,t_1)\;,
\end{equation}
is a necessary condition for $y(t)$ to be a Markov process for any $t_3\in(t_1,
t_2)$.[The opposite is not necessarily true, namely, if a stochastic process 
satisfies the Chapman-Kolmogorov equation, it is not necessarily Markov]. 
Therefore, one checks the validity of the Chapman-Kolmogorov equation for 
various values of $y_1$ by comparing the directly-evaluated $P(y_2,t_2|y_1,t_1)
$ with those computed according to right side of Eq. (13). 

The Markov time scale $t_M$ may be evaluated by the least-squares method. Since
for a Markov process one has
\begin{equation}
P(y_3,t_3|y_2,t_2;y_1,t_1)=P(y_3,t_3|y_2,t_2)\;,
\end{equation}
one compares $P(y_3,t_3;y_2,t_2;y_1,t_1)=P(y_3,t_3|y_2,t_2;y_1,t_1)P(y_2,t_2;
y_1,t_1)$ with that obtained based on the assumption of $y(t)$ being a Markov 
process. Using the properties of Markov processes and substituting in Eq. (14) 
yield
\begin{equation}
P_{\rm M}(y_3,t_3;y_2,t_2;y_1,t_1)=P(y_3,t_3|y_2,t_2)P(y_2,t_2;y_1,t_1)\;.
\end{equation}
One then computes the three-point joint probability distribution function 
through Eq. (14) and compares the result with that obtained through Eq. (15). 
Doing so entails, first, determining the quality of the fit by computing the 
least-squares fitting quantity $\chi^2$, defined by
\begin{equation}
\chi^2=\int dy_3\;dy_2\;dy_1\left[P(y_3,t_3;y_2,t_2;y_1,t_1)-
P_{\rm M}(y_3,t_3;y_2,t_2;y_1,t_1)\right]^2/(\sigma^2_{3j}+\sigma^2_{\rm M})\;,
\end{equation}
where $\sigma^2_{3j}$ and $\sigma^2_{\rm M}$ are, respectively, the variances 
of $P(y_3,t_3;y_2,t_2;y_1,t_1)$ and $P_{\rm M}(y_3,t_3;y_2,t_2;y_1, t_1)$.
Then, $t_M$ is estimated by the likelihood statistical analysis. In the absence
of a prior constraint, the probability of the set of three-point joint 
probability distribution functions is given by,
\begin{eqnarray}
& & P(t_3-t_1)=\cr\nonumber \\ & & \Pi_{y_3, y_2,y_1}
\frac{1}{\sqrt{2\pi(\sigma_{3j}^2+\sigma_{\rm M}^2)}}\exp\left\{
\frac{\left[P(y_3,t_3;y_2,t_2;y_1,t_1)-P_{\rm M}(y_3,t_3;y_2,t_2;
y_1,t_1)\right]^2}{2(\sigma_{3j}^2+\sigma_{\rm M}^2)}\right\}\;,
\end{eqnarray}
which must be normalized. Evidently then, when for a set of the parameters 
$\chi^2_\nu=\chi^2/N$ is minimum (with $N$ being the degree of freedom), the 
probability is maximum. Thus, if $\chi^2_\nu$ is plotted versus $t_3-t_2$, 
$t_M$ will be the value of $t_3-t_1$ at which $\chi^2_\nu$ is minimum [73]. 

Knowledge of $P(y_2,t_2|y_1,t_1)$ for a Markov process $y(t)$ is sufficient for
generating the entire statistics of $y(t)$, which is encoded in the $n-$point 
probability distribution function that satisfies a master equation, which 
itself is reformulated by a Kramers-Moyal expansion [74],
\begin{equation}
\frac{\partial P(y,t|y_0,t_0)}{\partial t}=\sum_k(-1)^k\frac{\partial^k}
{\partial y^k}\left[D^{(k)}(y,t)P(y,t|y_0,t_0)\right]\;.
\end{equation}
The Kramers-Moyal coefficients $D^{(k)}(y,t)$ are computed by,
\begin{eqnarray}
& & D^{(k)}(y,t)=\frac{1}{k!}\lim_{\Delta t\to 0}M^{(k)}\;,\nonumber\\
& & M^{(k)}=\frac{1}{\Delta t}\int dy'(y'-y)^kP(y',t+\Delta t|y,t)\;.
\end{eqnarray}
For a general stochastic process, all the coefficients can be nonzero. If, 
however, $D^{(4)}$ vanishes or is small compared to the first two coefficients 
[72], truncation of the Kramers-Moyal expansion after the second term is 
meaningful in the statistical sense, in which case the expansion is reduced to 
a Fokker-Planck equation that, in turn, according to the Ito calculus [72,74] 
is equivalent to a Langevin equation, given by
\begin{equation}
\frac{dy(t)}{dt}=D^{(1)}(y)+\sqrt{D^{(2)}(y)}\;\eta(t)\;,
\end{equation}
where $\eta(t)$ is a random ``force'' with zero mean and Gaussian statistics,
$\delta$-correlated in $t$, i.e., $\langle\eta(t)\eta(t')\rangle=2\delta(t-t')
$. 

The Langevin equation makes it possible to reconstruct a time series for $y(t)$
similar, {\it in the statistical sense}, to the original one, and can be used 
to make predictions for the future, i.e., given the state of the system at time
$t$, what would be the probability of finding the system in a particular state 
at time $t+\tau$. One writes $X(t+1)$ in terms of $X(t)$ by,
\begin{equation}
X(t+1)=X(t)\exp\{\sigma_y[y(t)+\bar{y}]\}\;,
\end{equation}
where $\bar{y}$ and $\sigma_y$ are the mean and standard deviations of $y(t)$.
To use Eq. (21) to predict $X(t+1)$, one needs $[X(t),y(t)]$. Thus, three
consecutive points in the series $y(t)$ are selected and a search is carried 
out for three consecutive points in the reconstructed $y(t)$ with the smallest 
difference with the selected points. Wherever this happens is taken to be the 
time $t$ which fixes $[X(t),y(t)]$. We now describe one application of the 
method.

{\it Example 1: fluctuations in human heartbeats.} It has been shown that 
various stages of sleep may be characterized by extended correlations of heart 
rates, separated by a large number of beats. The method described above based 
on the Markov time scale $t_M$ and the drift and diffusion coefficients, 
$D^{(1)}$ and $D^{(2)}$, provides crucial insights into the difference between 
the interbeat fluctuations of healthy subjects and patients with congestive 
heart failure. Figures 6 and 7 present [71,75] the drift and diffusion 
coefficients for the two groups of patients (for details of the data see the 
original references [71,72]). In particular, the diffusion coeffcients of the 
healthy subjects and those with congestive heart failure are completely 
different. Moreover, the important point to emphasize is that, the approach can
detect such differences even at the earliets stages of development of 
congestive heart failure [71,72], when no other analysis can.

Despite its success, the approach is still under development. According to the 
Pawula theorem [76], only three outcomes are possible in a Kramers-Moyal 
equation of order $k$: (a) The expansion is truncated at $k=1$, implying that 
the process is deterministic. (b) The expansion is truncated at $k=2$, which 
results in the Fokker-Planck equation describing a diffusion process, and (c) 
the expansion must, in principle, contain all the terms, $k\to\infty$, in which
case any truncation at a finite order $k>2$ would produce a non-positive 
probability distribution function, which is unphysical. More importantly, it 
has become evident [77] that a non-vanishing $D^{(4)}(X,t)$, i.e., if the 
Kramers-Moyal expansion cannot be truncated after the second term, represents 
a signature of a jump discontinuity in the time series, in which case one needs
the Kramers-Moyal coefficients of at least up to order six, i.e., up to 
$D^{(6)}(X,t)$, and in many cases even up to order eight [78], in order to 
estimate the jump amplitude and rate. For non-vanishing $D^{(4)}(X,t)$, the 
governing equation for a time series $X(t)$ with the jump-diffusion process is 
given by [77,78]
\begin{equation}
dX(t)=D^{(1)}(X,t)dt+\sqrt{D^{(2)}(X,t)}\;\eta(t)+\xi dJ(t)\;,
\end{equation}
where $J(t)$ is a Poisson jump process. The jump’s rate $\lambda(x,t)$ can be 
state-dependent with a size $\xi$, and is given by, $\lambda(x,t)=M^{(4)}(x,t)/
[3\sigma_\xi^4(x,t)]$, where, $\sigma^2_\xi(x,t)=M^{(6)}(x,t)/[5M^{(4)}(x,t)]$.
Dynamic processes with jumps are highly important, as they have been used to 
describe random evolution, for example, of neuron dynamics [79,80], soil 
moisture dynamics [81], and such financial features as stock prices, market 
indices, and interest rates [82], and epileptic brain dynamics [77]. Let us
describe a practical application of dynamic processes with jumps that is
data-based and reconstruct the governing equation for the dynamics.

{\it Example 2: reconstruction of stochastic dynamics of epileptic brain.} 
Brain's electrical rhythms in epileptic patients tend to become imbalanced, 
giving rise to recurrent seizures. When a seizure happens, the normal 
electrical pattern is disrupted by sudden and synchronized bursts of electrical
energy that may briefly affect the consciousness of the patient, as well as the
movements or sensations. Figure 8(b) presents intracranial 
electroencephalographic (iEEG) time series in a patient with seizures 
originating in the left mesial temporal lobe. 

The first- and second-order Kramers-Moyal coefficients and Langevin-type 
modelling of iEEG time series can be used to construct stochastic qualifiers 
of epileptic brain dynamics, which yield valuable information for diagnostic 
purposes. In particular, it has been shown [83] that qualifiers based on the
diffusion coefficient make it possible to obtain a more detailed 
characterization of spatial and temporal aspects of the epileptic process in 
the affected, as well as non-affected brain hemispheres. There is, however,
a major difference between the dynamics of the affected and non-affected 
regions of brain, with the former region, responsible for the generation
of focal epileptic seizures, being characterised by a non-vanishing 
fourth-order Kramers-Moyal coefficient, whereas that is not the case for the 
dynamics of latter region [83]. Thus, pathological brain dynamics is not 
described by the continuous diffusion processes and, hence, by the 
Langevin-type modelling described above. Pathological iEEG time series exhibit 
highly nonlinear properties [84], and are in fact described by a jump-diffusion
process.

Anvari {\it et al.} [77] considered intracranial iEEG time series, which
had been recorded during the pre-surgical evaluation of a subject with 
drug-resistant focal epilepsy. The multichannel recording [see Fig. 8(a)] 
lasted for about 2000s and was taken during the seizure-free interval from 
within the presumed epileptic focus (seizure-generating brain area), as well as
from distant brain regions. Thus, the analyzed data did not have a seizure 
event. Instead, it contained background iEEG time series. Anvari {\it et al.}
[77] showed that the $D^{(4)}$ coefficient of both time series do not vanish,
and modelled the data with a jump-diffusion process. Figure 9 presents the
computed diffusion coefficients $D^{(2)}(x)$, jump amplitudes 
$\sigma_\xi^2(x)$, and jump rates $\lambda(x)$, as well as the respective 
probability distribution functions, estimated from normalized iEEG time series 
that contained $4\times 10^5$ data points, for one epilepsy patient.

Carrying out extensive analyses of multi-day, multi-channel iEEG recordings 
from ten epilepsy patients, Anvari {\it et al.} [77] demonstrated that the 
dynamics of the epileptic focus is characterized by a stochastic process with 
a mean diffusion coefficient and a mean jump amplitude that are smaller than
those that characterize the dynamics of distant brain regions. Therefore,
higher-order Kramers-Moyal coefficients provide extra and highly valuable 
information for diagnostic purposes.

Note, however, that as a result of the jump processes, estimating the 
Kramers-Moyal coefficients by Eq. (16) encounters some fundamental drawbacks 
that have recently been studied [85-87]. Therefore, data-driven reconstruction 
of the governing equations based on Kramers-Moyal expansion is still an 
evolving approach, and as it is developed further, it will also find a wider 
range of applications.

\begin{center}
{\bf VI. DATA ASSIMILATION AND MACHINE LEARNING}
\end{center}

Even when we know the governing equations for a complex phenomenon, which are
in terms of ordinary or partial differential equations, and solve them 
numerically in order to describe the dynamic evolution of the phenomenon, 
uncertainties often remain and are usually of one of two types: (a) the 
internal variability that is driven by the sensitivity to the initial 
conditions, and (b) the errors generated by the model or the governing 
equations. The first type has to do with the amplification of the initial 
condition error, and arises even if the model is complete and ``perfect.'' It 
is mitigated by using data assimilation, briefly described in Sec III. The 
second type has recently been addressed by use of machine-learning techniques,
which have been emerging as an effective approach for addressing the issue of 
models' errors. As described above, in order to develop reduced-order models 
for complex phenomena, the variables and scales are grouped into unresolved and
resolved categories, and machine-learning approaches are emerging as being 
particularly suitable for addressing the errors caused by the unresolved 
scales. 

To see the need for addressing the errors due to unresolved scales, consider,
for example, the current climate models. The resolution of the computational 
grids used in the current climate models is around 50-100 km horizontally, 
whereas many of the atmosphere’s most important processes occur on scales much
smaller than such resolutions. Clouds, for example, can be only a few hundred 
meters wide, but they still play a crucial role in the Earth’s climate since 
they transport heat and moisture. Carrying out simulations at resolution is 
impractical for the foreseable future. Two approaches have been used to 
combine data assimilation with a machine-learning approach.

(i) The first approach is based on learning physical approximation, usually
called subgrid parameterization, which are typically computationally expensive.
Alternatively, the same can be achieved based on the differences between high- 
and low-resolution simulations. For climate models, for example, 
parametrizations have been heuristically developed over the past several 
decades and tuned to observations; see, for example, Hourdin {\it et al.} 
[88]. Due to the extreme complexity of the system, however, significant 
inaccuracies still persist in the parameterization, or physical approximations
of, for example, clouds in the climate models, particularly given the fact that
clouds also interact with such important processes as boundary-layer turbulence
and radiation. Given the debate over global warming and how much our planet 
will warm as a result of increased greenhouse gas concentrations, the fact that
such inaccuracies manifest themselves as model biases only goes to show the 
need for accurate and computationally affordable models. 

(ii) In the second approach one attempts to emulate the entire model by using 
observations, and spatially dense and noise-free data. Various types of
neural networks, including convolutional [89,90], recurrent [91], residual 
[92], and echo state networks [93] have been utilized. An echo state network is
a {\it reservoir computer} (i.e., a computational framework based on theory
of recurrent neural network that maps input data into higher-dimensional 
computational space through the dynamics of a fixed and nonlinear system called
a reservoir) that that uses a recurrent neural network with a hidden layer 
with low connectivity. The connectivity and weights of hidden neurons are fixed
and randomly assigned. Dedicated neural network architectures, combined with
a data assimilation method are used [94] in order to address problem of partial
and/or noisy observations.

As discussed by Rasp {\it et al.} [95], cloud-resolving models do alleviate 
many of the issues related to parameterized convection. Although such models
also involve their own tuning and parameterization, the advantages that they 
offer over coarser models are very significant. But climate-resolving models 
are also computationally too expensive, if one were to simulate climate change 
over tens of years in real time. Rapid increase in the computational power is 
making it possible, however, to carry out ``short'' time numerical simulations,
with highly resolved computational grids, that cover up to a few years. It is 
here that machine-learning approaches have begun to play an important role in 
addressing the issue of inaccuracies and grid resolution, because neural
networks can be trained by the results of the short-term simulations, and then 
be used for forecasting over longer periods of time.

{\it Example 1: Representing Subgrid Processes in Climate Models Using Machine 
Learning.} A good example is the approach developed by Rasp {\it et al.} [95] 
for representing subgrid processes in climate models. They trained a deep 
neural network to represent all atmospheric subgrid processes in a climate 
model. The training was done based on learning from a multiscale climate model 
that explicitly took into account convection. Then, instead of using the 
traditional subgrid parameterizations, the trained neural network was utilized 
in the global general circulation model, which could interact with the resolved
dynamics and other important aspects of the core model.

The base model that Rasp {\it et al.} utilized was version 3.0 of the 
well-known superparameterized Community Atmosphere Model (SPCAM) [96] in an
aquaplanet setup. Assuming a realistic equator-to-pole temperature gradient,
the sea temperature was held fixed, with a full diurnal cycle (a pattern that 
recurs every 24 hours), but no seasonal variation. In superparameterization, a
two-dimensional cloud-resolving model is embedded in each grid column (which in
Rasp {\it et al.}'s work was 84 km wide) of the global circulation model, which
resolves explicitly deep convective clouds and includes parameterizations for 
small-scale turbulence and cloud microphysics. For the sake of comparison, Rasp
{\it et al.} also carried out numerical simulations using a traditional 
parameterization package, usually referred to as the CTRLCAM. The model
and package exhibit many typical problems associated with traditional subgrid 
cloud parameterizations, including a double intertropical convergence zone,
and too much drizzle but also missing precipitation extremes, whereas SPCAM 
contains the essential advantages of full three-dimensional cloud-resolving 
models that address such issues with respect to observations.

The neural network used was a nine-layer deep, fully connected one with 256
nodes in each layer and $5\times 10^5$ parameters that were optimized in order 
to minimize the mean-squared error between the network’s predictions 
and the training targets. The advantages of the deep neural network are that 
they have lower training losses, and are more stable in the prognostic 
simulations. Simulations were carried out for five years, after a one-year 
spin-up (i.e.,  the time taken for an ocean model to reach a state of 
statistical equilibrium under the applied forcing). In the prognostic global 
simulations, the neural network parameterization interacted freely with the 
resolved dynamics, as well as with the surface flux scheme.

In Fig. 10(A) the results for the mean subgrig heating, computed by SPCAM, 
CTRLCAM, and neuralnetwork-aided model, referred to as NNCAM, are shown. The
results computed by the last two models are in very good agreement, whereas
those determined by simulating the CTRLCAM package produced a double peak, 
usually referred to as the intertropical convergence zone in climate model.
The corresponding mean temperatures are shown in Fig. 10(B), with the same
level of agreement between the resulyts based on SPCAM and NNCAM. The results
for the radiative fluxes predicted by the NNCAM parameterization are also in
close agreement with those of SPCAM for most of the globe, whereas CTRLCAM has
large differences in the tropics and subtropics caused by its aforementioned
double-peak bias. Figure 11 presents the results for precipitation 
distribution, indicating once again the inability of CTRLCAM in producing the
correct results, since the computed distribution exhibits too much drizzle
and absence of extremes. On the other hand, the results computed by SPCAM and
NNCAM are in good agreement, including the tails of the distribution.

In terms of speeding up the computations, NNCAM parameterization was about 20 
times faster than SPCAM’s. Moreover, the neural network does not become more 
expensive at prediction time, even if trained with higher-resolution training 
data, implying that the approach can scale with ease to neural networks trained
with much more expensive 3D global cloud-resolved simulations.

{\it Example 2: Inferring Unresolved Scale Parametrization of an 
Ocean-Atmosphere Model.} The second example that we briefly describe is the
work of Brajard {\it et al.} [97], who developed a two-step approach in which 
one trains model parametrization by using a machine-learning algorithm and 
direct data. Their approach is particularly suitable for cases in which the 
data are noisy, or the observations are sparse. In the first step a data 
assimilation technique was used, which was the ensemble Kalman filter, in order
to estimate the full state of the system based on a truncated model. The 
unresolved part of the truncated model was treated as model error in the data 
assimilation system. In the second step a neural network was used to emulate 
the unresolved part, a predictor of model error given the state of the system, 
after which the neural network-based parametrization model was added to the 
physical core truncated model to produce a hybrid model.

Brajard et al. [97] applied their approach to the Modular 
Arbitrary-Order-Ocean-Atmosphere Model (MAOOAM) [98], which has three layers,
two for the atmosphere and one for the ocean, and is a reduced-order 
quasi-geostrophic model that is resolved in the spectral space. The model
consists of $N_a$ modes of the atmospheric barotropic streamfunction 
$\psi_{a,i}$ and the atmospheric temperature anomaly $T_{a,i}$, plus $N_o$ 
modes of the oceanic streamfunction $\psi_{o,j}$ and the oceanic temperature 
anomaly $T_{o,j}$, so that the total number of variables is $N_x=2(N_a+N_o)$.
The ocean variables are considered as slow, while the atmospheric variables are
the fast ones. Two versions of MAOOAM were considered, namely, the true model 
with dimension $N_a=20$ and $N_o=8$ ($N_x=56$), and a truncated model with 
$N_a=10$ and $N_o=8$ ($N_x=36$). The latter model does not contain 20 
high-order atmospheric variables, ten each for the streamfunction and the 
temperature anomaly, and, therefore, it does not resolve the atmosphere-ocean 
coupling that is related to high-order atmospheric modes. 

The true model was used to simulate and generate synthetic data, part of which 
was used to train the neural network. The true model was simulated over 
approximately 62 years after a spin-up of 30,000 years. The synthetic
observations were slightly to take into account the fact that observations of 
the ocean are not at the same scale as those of the atmosphere; thus, before 
being assimilated, instantaneous ocean observations were averaged over a 55 
days rolling period centred at the analysis times. The architecture of the
neural network was a simple three layers multilayer perceptrons.

To test the accuracy and predictive power, as well as the long-term properties 
of the two versions of MAOOAM and their hybrid with a neural network, three
key variables, $\psi_{o,2}$, $T_{0,2}$ and $\psi_{a,1}$ - the second 
components of ocean streamfunction and temperature and the first component of 
the atmospheric streamfunction - were computed, since they account, 
respectively, for 42, 51, and 18 percent of the variability of the models.
Simulations of Brajard {\it et al.} [97] indicated that the predictions of the
hybrid model, one consisting of data assimilation and the neural network, with 
noisy data matched very closely with the hybrid model with perfect data. In
contrast, the truncated model's predictions differed from the true ones by
a factor of up to 3.

Wider application of the algorithm does face challenges. For example, the 
computational architecture, such as multi-core supercomputers and graphics 
processing units, and the data types used for physics-based numerical 
simulation and for machine-learning algorithms can be very different. Moreover,
training and running hybrid models efficiently impose very heavy requirements
on both the hardware and software. THese are, of course, challanges for an 
emerging field.

\begin{center}
{\bf VII. DATA-DRIVEN DISCOVERY OF THE GOVERNING EQUATIONS}
\end{center}

We now describe emerging approaches for discovering the governing equations
for complex phenomena in a complex system for which vast amounts of data may 
be available, but little is known about the governing equations for the 
physical phenomenon of interest, at least at the macroscale. Traditional 
approaches to analyzing such data rely on statistical methods and calculating 
various moments of the data, which in many cases are severly limited. There are
several emerging approaches to address this problem.

\begin{center}
{\bf A. Symbolic Regression}
\end{center}

While regression of numerical data and fitting them to an equation in order 
to better understand their implications is an old method, discovering the 
governing equation that describes the physics of a phenomenon for which data
are available, which are typically based on ordinary and partial differential
equations (ODEs and PDEs), involves manipulation of symbols and mathemathical
functions, such as derivatives and, therefore, represents a new type of 
regression. These methods, described in this and subsequent subsections, also
involve stochastic optimization for deriving the governing equations.

One of the first efforts for such systems was reported in the seminal papers of
Bongard and Lipson [99] and Schmidt and Lipson [100]. As Bongard and Lipson 
stated, ``A key challenge [to addressing the problem of having data but no 
governing equation], however, is to uncover the governing equations 
automatically merely by perturbing and then observing the system in intelligent
ways, just as a scientist would do in the presence of an experimental system. 
Obstacles to achieving this lay in the lack of efficient methods to search the 
space of symbolic equations and in assuming that pre-collected data are 
supplied to the modeling process.'' Since symbolic equations are typically in 
the form of ODEs and PDEs, the search space is quite large. 

Bongard and Lipson [99] described a method dubbed {\it symbolic regression}, 
which consists of three key elements: (a) {\it partitioning}, by which 
the governing equations that describe each of the system's variables are
synthesized separately, even though their behaviors may be coupled, hence 
reducing significantly the search space. (b) {\it Automated probing} that, in 
addition to modeling, automates (numerical) experimentation, leading to an {\it
automated scientific process}, and (c) {\it snipping}, which automatically 
simplifies and restructures models as they are synthesized to increase their 
accuracy, accelerate their evaluation, and make them more comprehensible for 
users. An automated scientific process tries [101] to mimic what many animals 
do, i.e., preserving the ability to operate after they are injured, by creating
qualitatively different compensatory behaviors 

In the symbolic regression algorithm [99,100] the partitioning is carried out 
by a stochastic optimization approach, of which there are many [102], such as 
simulated annealing [103] and the genetic algorithm [104]. Such methods are 
efficient enough for searching a relatively large space composed of building 
blocks, if the size of the dataset is not exceedingly large. Bongard and Lipson
[99] utilized the hill climbing method [105] for the optimization, a technique 
in which one begins optimization with an arbitrary solution, and then iterates 
it to generate a more accurate solution by making incremental changes to the 
last iterate. When the differential equation for variable $i$ is integrated 
numerically by, for example, a Runge-Kutta method, references to other 
variables are replaced by actual data. Bongard and Lipson [99] only tried to 
discover a set of first-order ODEs that governed the dynamics of the system 
that they studied. 

In symbolic regression approach a model consists of a set of nested expressions
in which each expression $i$ encodes the equations that describe the dynamic 
evolution of variable $i$. One also provides a set of possible mathematical 
operators, such as $\exp(\cdot)$, $\sin(\cdot)$, $d/dx$, etc., as well as 
operands that could be used to compose equations. During the first time step of
integrating the ODEs, each operand in each equation is set to be the initial 
conditions and the expression is evaluated, with the output being the 
derivative computed for that variable. The number of times that each model is 
integrated is the same as the number of times that the system has been supplied
with initial conditions, and all the models are optimized against all the time 
series observed or collected for the system.

There are at least three problems associated with symbolic regression. One is 
that it is computationally expensive since, in general, optimization typically 
requires intensive computations [102], unless certain ``tricks'' can be 
developed to accelerate them [102]. The second problem is the limitation of the
approach by the number of mathematical operations and their various 
combinations that it can carry out. The third shortcoming of the approach is 
that it could be prone to overfitting, unless one carefully balances model 
complexity with predictive power.

\begin{center}
{\bf B. Symbolic Regression and Genetic Programming} 
\end{center}

Improvements to the original symbolic regression approach are emerging. In 
particular, a genetic prgramming approacg, dubbed GPSR, which is a form of
symbolic regression, has emerged very recently that offers much promise. The 
GPs represent a kind of genetic algorithm in which models are represented as 
(nested) variable-length tree structures that represent a program, instead of 
a fixed-length list of operators and values.

We first recall that the genetic algorithm uses concepts from genetics and the 
Darwinian evolution to generate possible solutions for an optimization problem,
and involves four steps [102]: (a) {\it selection} for generating the 
solutions; (b) {\it design} of the ``genome'' to constrain the variables that 
define a possible solution, and the generation of the ``phenotype;'' (c) the 
{\it crossover} and {\it mutation} operations that are used for generating new 
approximate solutions and approaching the true optimal state, and (d) 
{\it elitism}, which selects the solutions that have the potential of 
eventually leading to the global optimal state. A {\it generation} of the 
computations is completed after the four sets of operations are carried out.

Similar to the theory of evolution according to which species that can adapt to
their environment produce the next generation of their offsprings - the updated
species - in an optimization problem solved by genetic algorithm each species, 
which is the set of all the parameters or, in the present problem, the model 
represented by an ODE or PDE that is to be discovered based on reproducing the 
given data, are selected by evaluating the cost function or, more generally, a 
{\it fitness function}, which is a measure of the quality and/or accuracy of 
the solution. Each possible solution is represented by a string of numbers, or 
``chromosomes,'' and after each round of testing or simulation, one deletes a 
number of the worst possible solutions, and generates new ones from the best 
possible solutions. Therefore, a figure of merit or fitness is attributed to
each possible solution that measures how close it has come to meeting the 
overall specification. This is done by applying the fitness function to the 
simulation results obtained from that possible solution. The species with a 
smaller cost function, or better fitness, has a higher probability of producing
one or more offsprings, i.e., possibly more accurate solutions in the form of 
ODEs or PDEs, for the next generation, which is usually referred to as the {\it
population}. 

Using the population of the species, one solves the proposed ODE or PDE, 
computes the properties for which data are given, and evaluates the cost or
fitness function, in order to choose the ODE or PDE that is more likely to 
produce more accurate, next generation predictions for the data. Such 
candidates are randomly recombined - the crossover step - and permuted - the 
mutation step - to generate new candidate equations. The candidates with the 
highest cost function, or the poorest fitness, are eliminated from the 
population, a step that represents natural selection in Darwinian evolution.

An illuminating example is a very recent application of GPSR [106] to anomalous
diffusion [107] in the incipient percolation cluster at the percolation 
threshold [108,109], which is a fractal (and macroscopicaaly heterogeneous) 
structure at all the length scales with a fractal dimension $D_f$ whose values 
in 2D and 3D are, respectively, $91/48\simeq 1.9$ and 2.53. Diffusion in the 
cluster is anomalous [107], i.e., the mean-squared displacement of a diffusing 
particle grows with time as, $\langle R^2(t)\rangle\propto t^\alpha$, where 
$\alpha=2/D_w$, with $D_w$ being the fractal dimension of the walk with, 
$D_w\simeq 2.87$ and 3.8 in 2D and 3D. An important, and for quite sometime 
controversial, issue was the governing equation for $P({\bf r},t)$, the average
probability that a diffusing particle is at position {\bf r} at time $t$, for 
which various equations [110-112] were suggested. 

Using numerical simulation of diffusion on the incipient percolation cluster 
in 2D by random walks, Im {\it et al.} [106] collected extensive numerical data
for $P({\bf r},t)$. When they applied the GPSR method to the data, they 
discovered that the governing equation for $P({\bf r},t)$ is given by
\begin{equation}
\frac{\partial^{0.62}P}{\partial t^{0.62}}=\frac{0.82}{r}\frac{\partial P}{
\partial r}+\frac{\partial^2P}{\partial r^2}\;.
\end{equation}
where $\partial^\alpha/\partial t^\alpha$ indicates fractional derivative. Note
that the factor $1/r$ in the first term of the right side of Eq. (23) was 
discovered by the algorithm, and was not included in the set of trial searches.
The governing equation for $P(r,t)$, derived by Metzler {\it et al.} [112], is 
given by
\begin{equation}
\frac{\partial^\alpha P}{\partial t^\alpha}=\frac{1}{r^{d_s-1}}\frac{\partial}
{\partial r}\left[r^{d_s-1}\frac{\partial P(r,t)}{\partial r}\right]=\frac{
d_s-1}{r}\frac{\partial P}{\partial r}+\frac{\partial^2P}{\partial r^2}\;,
\end{equation}
where $d_s=2D_f/D_w$, with $\alpha\approx 0.7$. Thus, the discovered equation 
and one that is generally accepted to govern anomalous diffusion in the 
incipient percolation cluster at the percolation threshold are practically 
idential.

He {\it et al.} [113] showed that the dynamics of transport processes in 
heterogeneous media that are described by a fractional diffusion equation is 
not self-averaging, in that time and ensemble averages of the observables, such
the mean-squared displacements, do not converge to each other. This is 
consistent with what is known for diffusion on the CPC at the percolation 
threshold [114,115], for which the distribution of the displacements of the 
diffusing particle does not exhibit self-averaging. The discovery of a 
fractional diffusion equation for diffusion on the critical percolation
cluster at the percolation threshold is fully consistent with this picture, and
indicates the internal consistency accuracy of the approach.

The GPSR has also been used to discover morphology-dependent plasticity models 
for additively-manufactured Inconel 718 [116]. Although the genetic algorithm 
is amenable to parallel processing and computations, the GPSR, at this point, 
is not, since it involves numerically solving a population of ODEs or PDEs. 
Thus, one needs to develop more efficient ways of solving them in order to turn
GPSR into a powerful and reliable tool for large-scale scientific problems. 

\begin{center}
{\bf C. Sparse Identification of Nonlinear Dynamics}
\end{center}

Schmid [117] proposed the dynamic mode decomposition method, a dimensionality 
reduction algorithm for time series in fluid systems. The algorithm, an 
effective method for capturing the essential features of numerical or 
experimental data for a flow field, computes a set of modes, each of which is 
associated with a fixed oscillation frequency and decay/growth rate, and 
represent approximations of the modes and eigenvalues of the composition 
operator, which is also referred to as the Koopman operator [118]. Jovanovi\'c 
{\it et al.} [119] developed a sparsity-promoting variant of the original 
dynamic mode decomposition algorithm in which sparsity was induced by 
regularizing the least-squared differences between the matrix of snapshots of 
a system and a linear combination of the modes, with an additional term that 
penalizes the L1-norm - the sum of the magnitudes of the vectors in a space - 
of the vector of dynamic mode decomposition amplitudes. As the name suggests, 
the only assumption of the algorithm about the structure of the model is that, 
there are only a few important terms that govern the dynamics of a system, 
implying that the searched-for equations are sparse in the space of possible 
functions, an assumption that holds for many physical systems. 

As an important improvement and extension to the original symbolic regression 
algorithm, Brunton {\it et al.} [120] proposed a method, the sparse 
identification of nonlinear dynamics (SINDy). Sparse regression, used for 
discovering the fewest terms in the governing equations that are required for 
accurately representing the data, avoids overfitting that often occurs in such 
approches. Brunton {\it et al.} [120] considered dynamical systems of the type,
\begin{equation}
\frac{d{\bf x}(t)}{dt}={\bf f}[{\bf x}(t)]\;,
\end{equation}
in which ${\bf x}(t)$ is the state of the system at time $t$, and ${\bf f}[{\bf
x}(t)]$ represents the dynamic constraints that define the equations of motion 
of the system, such as, for example, the Navier-Stokes equations for 
hydrodynamics of Newtonian fluids, which can be generalized to include
parameterization and forcing. Thus, the goal is to determine ${\bf f}[{\bf x}
(t)]$ from the data.

To do so, one collects [120] a time history of the state ${\bf x}(t)$ and 
either measures the derivative $d{\bf x}(t)/dt=\dot{\bf x}(t)$ or approximates
it numerically. The data are then sampled at several times $t_1$, $t_2$, 
$\cdots$, $t_m$ and organized into two matrices, given by
\begin{equation}
{\bf X}=\left[\begin{array}{cccc}
x_1(t_1) & x_2(t_1) & \cdots & x_n(t_1)\\
x_1(t_2) & x_2(t_2) & \cdots & x_n(t_2)\\
\vdots & \vdots & \vdots & \vdots\\
x_1(t_m) & x_2(t_m) & \cdots & x_n(t_m)
\end{array}\right]\;,
\end{equation}
and
\begin{equation}
\dot{{\bf X}}=\left[\begin{array}{cccc}
\dot{x}_1(t_1) & \dot{x}_2(t_1) & \cdots & \dot{x}_n(t_1)\\
\dot{x}_1(t_2) & \dot{x}_2(t_2) & \cdots & \dot{x}_n(t_2)\\
\vdots & \vdots & \vdots & \vdots\\
\dot{x}_1(t_m) & \dot{x}_2(t_m) & \cdots & \dot{x}_n(t_m)
\end{array}\right]\;.
\end{equation}
One then sets up a library ${\cal L}({\bf X})$ of candidate nonlinear functions
of the columns of {\bf X}, with each column representing a candidate function 
for the right side of Eq. (25). There is, of course, complete freedom in 
selecting the candidate functions. For example,
\begin{equation}
{\cal L}({\bf X})=\left[1\;\;{\bf X}\;\;{\bf X}^{(2)}\;\;{\bf X}^{(3)}\;\cdots
\; \sin({\bf X})\;\;\cos({\bf X})\; \cdots\right]\;,
\end{equation}
where ${\bf X}^{(n)}$ denotes a polynomial of order $n$. Thus, for example,
\begin{equation}
{\bf X}^{(2)}=\left[\begin{array}{cccccc}
x_1^2(t_1) & x_1(t_1)x_2(t_1) & \cdots & x_2^2(t_1) & \cdots & x^2_n(t_1)\\
x_1^2(t_2) & x_1(t_2)x_2(t_2) & \cdots & x_2^2(t_2) & \cdots & x^2_n(t_2)\\
\vdots & \vdots & \vdots & \vdots & \vdots & \vdots\\
x_1^2(t_m) & x_1(t_m)x_2(t_m) & \cdots & x_2^2(t_m) & \cdots & x^2_n(t_m)
\end{array}\right]\;.
\end{equation}
If we know, for example, that only a few of the nonlinearities are active in 
each row of the ${\bf f}({\bf x})$ in Eq. (25), we set up a sparse regression 
problem to determine the sparse vectors of coefficients, ${\bf\Xi}=\left[\mbox
{\boldmath$\chi$}_1,\mbox{\boldmath$\chi$}_2\cdots\mbox{\boldmath$\chi$}_n
\right]$ that determine which nonlinearities are active:
\begin{equation}
\dot{{\bf X}}={\cal L}({\bf X}){\bf\Xi}\;.
\end{equation}
Each column of $\mbox{\boldmath$\chi$}_k$ is a sparse vector of coefficients
that determines the terms that are active on the right side of Eq. (25). After
${\bf\Xi}$ is determined, a model for each row of the governing equations is 
constructed by
\begin{equation}
\frac{d{\bf x}}{dt}={\bf f}_k({\bf x})={\cal L}({\bf x}^{\rm T})
\mbox{\boldmath$\chi$}_k\;.
\end{equation}
In Eq. (31) ${\cal L}({\bf x}^{\rm T})$ is a vector of symbolic functions of
elements of {\bf x}, whereas ${\cal L}({\bf X})$ is the data matrix. In other
words
\begin{equation}
\frac{d{\bf x}}{dt}={\bf f}({\bf x})={\bf\Xi}^{\rm T}\left[{\cal L}({\bf x}^{
\rm T})\right]^{\rm T}\;.
\end{equation}

Note that each column of Eq. (30) requires a separate optimization to determine
the sparse vector $\mbox{\boldmath$\chi$}_k$ for the $k$th-row equation. In 
general, ${\cal L}({\bf X})$ is a $m\times p$ matrix, with $p$ being the number
of candidate functions. Naturally, $m\gg p$ because, typically, there are far 
more data than functions. Since the number of functional forms can be very 
large, one tests many different function bases and uses the sparsity and 
accuracy of the resulting model as a way of determining the correct basis to 
represent the data. The testing can be guided by knowledge about the physics
of the problem.

It should be clear that the success of the application of the method to any
phenomena and the accuracy of the resulting model depend on the choice of 
measurement variables, quality of the data, and the sparsifying function basis.
While it may be difficult to know the correct variables a priori, time-delay 
coordinates often provide useful variables from a time series [121,122]. In 
this method, vectors in a new space, referred to as the embedding space, are 
formed from time-delayed values of the measurements,
\begin{equation}
s_m=\left[s_{n-(d-1)\tau},s_{n-(d-2)\tau},\cdots,s_n\right]\;,
\end{equation}
where $d$ is the embedding dimension, and $\tau$ is the time lag or delay.
According to Takens [121], if a sequence $\{s_m\}$ consists of measurements of 
the state of a dynamical system, then, under certain generic assumptions, the 
time-delay embedding provides a one-to-one image of the original set, if $d$ is
large enough. If one has $m$ available measurements, the number of embedding 
vectors is only $m-(d-1)\tau$. Of course, knowledge about the physics of the
phenomena of interest also helps one to identify reasonable choices of 
nonlinear functions and measurement coordinates. For example, problems in 
hydrodynamics have to do with the momentum conservation equations, and for 
Newtonian fluids with the Navier-Stokes equations.

For many important problems in science and engineering, such as those in 
hydrodynamics and transport and deformation in heterogeneous materials, the 
phenomena of interest are represented by PDEs that contain a few spatial 
variables, and involve either a very large number of measured data, or 
numerical data obtained from micro-scale simulations. Straightforward 
application of the method to such problems will be impractical, since the 
factorially growth of the library ${\cal L}$ with $m$ in Eq. (29) and the 
required number of separate optimizations make such applications impractical. 
But a solution has also been developed. Consider, for example, a fluid flow 
problem in 3D space, governed by the Navier-Stokes equations. One can use the 
proper orthogonal decomposition technique [123] that reduces the complexity of 
intensive numerical simulations that, in the present context, implies that the 
Navier-Stokes equations are replaced by simpler models that require much less 
computations to solve numerically; see also the above example for modeling the
artery system in human body.

{\it Example: Vortex Shedding Behind a Cylinder.} An illuuminating application 
of the SINDy was made by Brunton {\it et al.} [120] to the classical problem of
vortex shedding behind a cylinder. It was suggested a long time ago [124] that 
turbulent flow arises as a result of a series of Hopf bifurcations, 
representing cubic nonlinearities. Such nonlinearity was puzzling because the 
Navier-Stokes equations contain only quadratic nonlinearity (it is a 
second-order PDE). When the first Hopf bifurcation was actually discovered 
[125,126] during the transition from a steady laminar wake to laminar periodic 
vortex shedding at Reynolds number, Re $=47$, it was shown [127] that a 
coupling between oscillatory modes and the base flow gives rise to a slow 
manifold that results in algebraic terms that approximate cubic nonlinearities 
on slow time scales. 

Using data obtained by numerical simulation of the Navier-Stokes equations past
a cylinder at a Reynolds number Re $=100$ reported by Colonius and Taira [128],
Brunton {\it et al.} [120] showed that their approach recovers the Hopf normal 
form, a problem that had taken 30 years to resolve. Since the Navier-Stokes 
equations contain quadratic nonlinearity, Brunton {\it et al.} had to use a 
mean-field model with a separation of time scales, such that a fast mean-field
deformation was slave to the slow vortex shedding dynamics. Thus, they used a 
reduced-order mean-field model for the cylinder dynamics, proposed by Noack 
{\it et al.} [127],
\begin{eqnarray}
\frac{dx}{dt} & = & \mu x-\omega y+Axz\;,\\
\frac{dy}{dt} & = & \omega x+\mu y+Ayz\;,\\
\frac{dz}{dt} & = & -\lambda(z-x^2-y^2)\;.
\end{eqnarray}
For large values of $\lambda$, the $z$ dynamics would be slow and, therefore, 
the mean flow would rapidly correct and be on the slow manifold, $z=x^2+y^2$, 
given by the aplitude of vortex shedding. The Hopf normal form is recovered by 
substituting the algebraic forms into Eqs. (34) and (35).

Given the time history of the three coordinates, the SINDy algorithm correctly 
identified quadratic nonlinearities (in the Navier-Stokes equations) and 
reproduced a parabolic slow manifold. Equations (34) - (36) involve the 
derivatives whose measurements were not available, but were computed from the 
state variables. More importantly, when the training data do not include 
trajectories that originate off of the slow manifold, the algorithm 
{\it incorrectly} identifies cubic nonlinearities, hence failing to identify 
the slow manifold.

Figure 12 presents the results and compares them with full simulations. The 
parabolic slow manifold is shown on the left side of Fig. 12, which contains 
vortex shedding indicated by A, the mean flow indicated by B, and an unstable 
fixed point C. A proper orthogonal decomposition basis and shift mode were used
in order to reduce the dimension of the problem, shown in the middle right of 
the figure. The agreement between the identified dynamics and the true
trajectory in the proper orthogonal decomposition coordinates is excellent. The
identified dynamics also captures the quadratic nonlinearity and time scales 
associated with the mean-field model.

The open source software package [129] PySINDy [Pyton SINDy] has been developed
in Python to integrate the various versions of SINDy [130]. Note that by 
promoting sparsity, SINDy solves an over-determined set of equations, ${\bf Ax}
={\bf b}$, making it modular and, hence, amenable to innovations. Compared with
the original symbolic regression described above, SINDy is extremely efficient 
computationally, requiring orders of magnitude less computation time. It may 
also be used with neural networks that provide automatic differentiation 
[131,132], and learning coordinates and models jointly [133,134]. Even though
the approach has been applied to a wide variety of problems [135-156] over the 
past few years, it is still evolving in order to make it applicable to a wider 
class of problems, as well as making it faster computationally.

A distinct version of SINDy, weak sparse identification of nonlinear dynamics 
(WSINDy), first proposed by Schaeffer and McCalla [156] and improved 
significantly by Messenger and Bortz [157], attempts to bypass computations of 
the derivatives required by SINDy, hence increasing significantly the speed of 
the computations. The approach assumes that the function ${\bf f}({\bf x})$ in 
Eq. (25) can be accurately represented by polynomials, $F(x)=x^{j-1}$, and 
utilizes a number of feature vectors that are large enough to include all the 
terms present in the underlying system. Each feature vector $v_j(x,t_k)$ is 
approximated by using piecewise constant quadrature,
\begin{equation}
v_j(x,t_k)=\int_0^{t_k}F_j[x(t)]dt\approx\Delta t\sum_{l=1}^k F_j[x(t_l)]\;,
\end{equation}
with $k=1,2,\cdots,K$, and $v_j(x,t_0)=v_j(x,0)=0$, and $K$ and $\Delta t$ 
being, respectively, the number of discrete time steps, and the size of time 
steps. The quadrature yields a close approximation to the noiseless ${\bf x}
(t)$ without smoothing, and effectively calculates a scaled expectation $E$ for
a sum of random variables of the form, $x^n\eta^p$, $E(x^n\eta^p)=E(x^n)E
(\eta^p)$. Decoupling of the two expected values is permitted since the noise 
is sampled independently of the data. Thus, many of the noise-dependent cross 
terms are essentially zero, if piecewise constant quadrature is used to 
approximate the feature vector.

By eliminating pointwise derivative approximations, one obtains estimates for 
the model's coefficients from noise-free data with machine precision, as well 
as robust identification of PDEs with large noise in the data. One discretizes 
a convolutional weak form of the PDE, and utilizes separability of the test 
functions for efficient model identification using fast Fourier transform. 
Messenger and Bortz [158] showed that WSINDy algorithm has, at worse, a 
computational complexity on the order of ${\cal O}(N^{d+1}\log N)$ for $N$ data
points in each of $d+1$ dimensions, i.e., ${\cal O}(\log N)$ operations per 
datapoint. The approach has been used to study a number of important problems 
involving complex phenomena [159-162]. 

\begin{center}
{\bf D. The Mori-Zwanzig Formulation}
\end{center}

Mori [163] and Zwanzig [164] developed a formalism that provides a 
mathematically exact procedure for developing reduced-order models for 
high-dimensional dynamical systems, such as turbulent flow, as well as data, 
which are constructed based on projection operators. The essence of the method 
is reformulating a set of ODEs into a reduced system for the resolved variables
$x_r$, but still retaining the dynamics of the original system, which implies 
correctly representing the contribution of the unresolved variable on the 
resolved physics of the system. It does so by applying a projection operator to
the evolution process of the original dynamic systems described by the set of 
ODEs, in order to chieve reduction in their dimensionality. 

Mori's formulation leads to a generalized linear Langevin equation, whereas 
that of Zwanzig produces generalized nonlinear Langevin equation. The
equation consists of Markovian, noise, and memory terms, and is an {\it exact} 
representation of the dynamics of the model. Thus, the approach may be viewed 
as a nonlinear generalization of the stochastic Kramers-Moyal expansion, 
described above, in the limit that only the first two terms of the expansion 
are important, since in that limit one obtains a description of the system by a
linear Langevin equation. In practice, however, use of the method is 
computationally difficult, particularly when applied to systems that are 
described by PDEs; this is discussed below. Comprehensive discussions of the 
subject are given Mazenko [165], Evans and Morriss [166], and Hi\'jon {\it et 
al.}  [167].

Before describing the Mori-Zwanzig approach, let us point out that the 
procedure was originally developed for describing non-equilibrium statistical 
mechanics of molecular systems, with the goal of solving for the probability 
density functions and time correlation functions of non-equilibrium systems, 
and was limited to Hamiltonian dynamical systems. Chorin {\it et al.} [168] 
extended the formulation to general time-dependent systems, such as those in 
hydrodynamics and reaction-diffusion systems. They developed their framework 
for optimal prediction, i.e., obtaining the solution of nonlinear 
time-dependent problems, described by Eq. (25), for which a full-order solution
is too difficult computationally and, in addition, the unresolved part of the 
initial conditions is uncertain.

We describe the Mori-Zwanzig formulation by closely following Falkena {\it et 
al.} [169]. Consider nonlinear dynamical systems described by Eq. (25).
Consider an initial condition corresponding to a trajectory $x(t)$, $x(t=0)=y$,
and an observable $u(y,t)=g[x(t)]$ along a solution of Eq. (25), where $g$ is 
defined on $R^n$. Thus, one must have
\begin{equation}
\frac{\partial}{\partial t}u(y,t)={\cal L}u(y,t)\;,
\end{equation}
with $u(y,0)=g(y)$, where ${\cal L}$ is the Liouville operator defined by, 
${\cal L}u=\sum_{i=1}^nR_i(y)\partial u(y,t)/\partial y_i$, with $y_i$ being 
the $i$th component of $y$, and {\bf R} the vector field of Eq. (25). The goal 
for a linear system is to construct a system of equations for a select subset 
of $m$ resolved variables $x_r\in R^m$, with the unresolved variables denoted 
by $x_u\in R^{n-m}$, such that, ${\bf x}=(x_r,x_u)$. 

To reduce or map the system of $n$ components to one with $m$ components, one 
needs a projection operator $P$, $P:C(R^n,R^k)\to C(R^m, R^k)$, with $k$ being 
the dimension of an arbitrary function $f(x_r,x_u)$ to which the projection is 
applied. One example of such projection operator is the linear one, 
$[Pf](x_r,x_u)=f(x_r,0)$, i.e., one that sets all the unresolved variables to 
zero, keeping only the resolved ones. We denote by $Q$ the complement of $P$, 
defined by, $Q=I-P$, where $I$ is the identity operator.

The solution of the system (38) is, $u(y,t)=[e^{t{\cal L}}g](y)$, where 
$e^{t{\cal L}}$ is the evolution, or the aforementioned Koopman operature 
[118], which propagates an observable with ${\cal L}$ and $e^{t{\cal L}}$ 
commuting. Thus, if $g=y_i$, then, $x_i(y,t)=e^{t{\cal L}}y_i$. Equation (38) 
is rewritten as
\begin{equation}
\frac{\partial}{\partial t}\left[e^{t{\cal L}}g\right](y)=\left[e^{t{\cal L}}
{\cal L}g\right](y)=\left[e^{t{\cal L}}P{\cal L}g\right](y)+\left[e^{t{\cal L}}
Q{\cal L}g\right](y)
\end{equation}
The second term on the right side of Eq. (39) describes the evolution of the
unresolved variables. If one invokes the Duhamel-Dyson identity, namely,
\begin{equation}
e^{t(A+B)}=e^{tA}+\int_0^t e^{(t-s)(A+B)}Be^{sA}ds\;,
\end{equation}
and take, $A=Q{\cal L}$ and $B=P{\cal L}$, one obtains
\begin{equation}
\left[e^{t{\cal L}}Q{\cal L}g\right](y)=\left[e^{tQ{\cal L}}Q{\cal L}g\right]
(y)+\int_0^t\left[e^{(t-s){\cal L}}P{\cal L}e^{sQ{\cal L}}Q{\cal L}g\right](y)
ds\;,
\end{equation}
and, therefore, Eq. (39) becomes
\begin{equation}
\frac{\partial}{\partial t}\left[e^{t{\cal L}}g\right](y)=\left[e^{t{\cal L}}
P{\cal L}g\right](y)+\left[e^{tQ{\cal L}}Q{\cal L}g\right](y)+\int_0^t\left
[e^{(t-s){\cal L}}P{\cal L}e^{sQ{\cal L}}Q{\cal L}g\right](y)ds\;.
\end{equation}
In particular, if $g(y)=y_i$, we have, $[e^{t{\cal L}}g](y)=x_i(y,t)$, and 
obtain the generalized Langevin equation,
\begin{equation}
\frac{\partial}{\partial t}x_i(y,t)=M_i[x_r(y,t),0]+{\cal N}_i(y,t)+\int_0^t
K_i[x_r(y,t-s),s]ds\;,
\end{equation}
where, ${\cal N}_i=\left[e^{tQ{\cal L}}Q{\cal L}g\right](y)$, and, $K_i=
[P{\cal L}{\cal N}_i](y,t)$. As mentioned above, the Mori-Zwanzig formulation 
produces a generalized Langevin equation with three terms, namely, the Markov, 
noise, and memory functions represented, respectively, by $M_i$, ${\cal N}_i$, 
and the integral on the right side of Eq. (43). The noise term, produced by the
uncertainty in the initial conditions, is the solution of the following 
orthogonal dynamic equation,
\begin{equation}
\frac{\partial}{\partial t}{\cal N}_i(y,t)=Q{\cal L}{\cal N}_i(y,t)\;,
\end{equation}
with the initial condition, ${\cal N}_i(y,0)=Q{\cal L}y_i$. It is called 
orthogonal dynamics because its solution lies in the orthogonal space of 
projection operator $P$ at all times.

Equation (43) is exact, but determining its solution is not necessarily simpler
than the original equation, Eq. (25). The main bottleneck for using the 
Mori-Zwanzig approach to construct dynamical equations for a set of data, i.e.,
using Eq. (43), is determining the numerical solution of Eq. (44), which is 
difficult. For example, directly evaluating the integral requires storing the 
solutions from all previous steps at every time step, which is a difficult 
task. The ``ease'' of obtaining the solution depends crucially on the choice of
projection operator $P$, which plays an important role in determining the form 
and complexity of the orthogonal dynamics equation, Eq. (44). $P$ should be 
selected such that the orthogonal dynamics system is stable, implying that one
must not only retain stabilizing factors in the unresolved dynamics, but also
select $P$ such that solving Eqs. (43) and (44) is less complex than solving
the original system described by Eq. (25). 

A simple example [169,170] illustrates how the method works. Consider the 
following system of ODEs,
\begin{equation}
\frac{d}{dt}\left(\begin{array}{c}
x_r\\
x_u
\end{array}\right)=\left(\begin{array}{cc}
a_{11} & a_{12}\\
a_{21} & a_{22}  
\end{array}\right)
\end{equation}
with the initial condition, $x(0)=(y_r,y_u)$. We wish to derive an equation for
$x_r$ only, which is accomplished by solving the equation for $x_u$ by the
method of variation of constants and substituting the result into the equation 
for $x_r$ to obtain
\begin{equation}
\frac{dx_r}{dt}=a_{11}x_r(t)+a_{12}e^{a_{22}t}y_u+\int_0^t a_{12}e^{a_{22}
(t-s)}a_{21}x_r(s)ds\;.
\end{equation}
Equation (46), which is exact, exhibits the same behavior as the original 
system, and the effect of the unresolved variables appears only as the initial 
condition $y_u$. 

Since the main obstacle to using the Mori-Zwanzig approach is having the right 
projection operator $P$, it may be useful to discuss the issue further, so as
to provide some guidance for selecting the operator. As already discussed, 
Mori's formulation leads to a linear generalized Langevin equation, whereas 
that of Zwanzig produces nonlinear generalized Langevin equation. In the former
case, the projection operator relies on the inner product defined defined by, 
$\langle f,g\rangle=\int f({\bf x})g({\bf x})d\mu({\bf x})$, where $\mu({\bf x}
)$ is the probability distribution function. Given the inner product, Mori’s 
projection operator is defined onto the span of a set of linearly independent 
basis functions $b_i({\bf x})$, so that [171]
\begin{equation}
[Pf]{\bf b}({\bf x})=\sum_i\sum_j\langle f,b_i\rangle[{\cal M}^{-1}]_{i,j}b_j
({\bf x})\;,
\end{equation}
with ${\cal M}$ being the covariance matrix, ${\cal M}_{ij}=\langle b_i,b_j
\rangle$. If the basis functions are orthonormal, then, ${\cal M}={\bf I}$, 
with {\bf I} being the identity matrix, and the projection operator is greatly 
simplified:
\begin{equation}
[Pf]{\bf b}({\bf x})=\sum_o\langle f,b_i\rangle b_i({\bf x})\;.
\end{equation}

On the other hand, in Zwanzig's formulation, the observables are a subset of 
the resolved variables ${\bf x}_r$, and the projection operator is defined by 
direct marginalization of the unresolved variables. If the probability 
distribution $\mu({\bf x})$ for variable {\bf x} is written for 
resolved/unresolved variables as a density function $\rho({\bf x}_r,{\bf 
x}_u)$ [171], then,
\begin{equation}
[Pf]({\bf x}_r)=\frac{\displaystyle\int f({\bf x}_r,{\bf x}_u)\rho({\bf x}_r,
{\bf x}_u)d{\bf x}_u}{\displaystyle\int\rho({\bf x}_r,{\bf x}_u)d{\bf x}_u}
\end{equation}
which yields a nonlinear function that has been used [172] for developing 
models of turbulence based on the Mori-Zwanzig formulation.

Alternative ways of getting around the difficulty of selecting the projection
operator have also been suggested. For example, Gouasmi {\it et al.} [170] 
proposed to approximate the orthogonal dynamics equation by a less complex one
using pseudo-orthogonal dynamics approximation. In their method the memory 
kernel in the above integral is estimated a priori by utilizing full-order 
solution snapshots. Thus, a pseudo-orthogonal dynamics equation is solved that 
has the Liouville form, instead of solving the original one. The method is 
based on the assumption that, for one observable, the semi-group of the 
orthogonal dynamics operator is a composition operator, akin to semi-groups of 
Liouville operators, hence mimicking their behavior. 

Despite the difficulty in developing the right projection operator $P$ and 
obtaining a numerical solution for the dynamics of the system that is less
expensive than solving the original system, the Mori-Zwanzig approach is
gradually gaining more recognition and use as a way of discovering the
governing equations for systyems for which a considerable amount of data is 
available. Chu and Li [173] used the procedure to derive an equation that 
describes heat conduction in nano-mechanical systems, since the conventional 
heat conduction equation breaks down at such length scales. They considered a 
1D isolated chain of $N$ atoms, divided evenly into $n$ blocks, each of which 
contained ``atoms'' with known equilibrium spacing between two atoms, and 
calculated energy transport between the blocks. Thus, the local energy density 
was selected as the coarse-grained variable, for which a generalized Langevin 
equation was derived using the Mori-Zwanzig procedure. The propagating operator
${\cal L}$ was defined by
\begin{equation}
{\cal L}\equiv v_0\frac{\partial}{\partial x_0}+\frac{f(x_0)}{m}\frac{\partial}
{\partial v_0}
\end{equation}
where $x_0$ and $v_0$ are, respectively, the initial position and velocity of 
the molecules, $m$ is their mass, and $f(x)$ is the force, i.e., $f=-\mbox
{\boldmath$\nabla$}E(x)$, with $E$ being the potential energy. They showed that
the calculated result with the Mori-Zwanzig method agrees with the results of 
non-equilibrium molecular dynamics simulations in which they imposed a
temperature gradient between the two ends of the chain of 250 atoms.

{\it Example: Reduced-Order Equation for Turbulent Flow.} Tian {\it et al.} 
[171] used extensive data for isotropic turbulence in order to drive the 
projection operator of Mori-Zwanzig approach and construct a reduced-order 
Navier-Stokes equation. If the Navier-Stokes equation is spatially discretized,
one obtains the following set of nonlinear equations for the fluid's velocity 
${\bf v}(t)$:
\begin{equation}
\frac{d{\bf v}(t)}{dt}=R[{\bf v}(t)]\;,
\end{equation}
which is of the form given by Eq. (25), where $R$ is the nonlinear function 
that represents the spatially discretized right side of the Navier-Stokes 
equations. Computations for fully resolved dynamics of the Navier-Stokes 
equations is prohibitive for any physical problem. Thus, to develop a 
reduced-order model for turbulence, the velocity field is usually 
coarse-grained using a spatial filter, which reduces the range of scales that 
must be resolved. Suppose that $\bar{{\bf v}}(t)$ is the filtered fluid 
velocity. Then, as described above, according to the Mori-Zwanzig formulation, 
Eq. (43) for $\bar{{\bf v}}(t)$, in vector form, is given by
\begin{equation}
\frac{d\bar{{\bf v}}(t)}{dt}=M[\bar{{\bf v}}(t)]+{\cal N}(t)-\int_0^t 
K[\bar{{\bf v}}(t-s),s]ds\;,
\end{equation}
which is the nonlinear version of the formulation. In the linear formulation,
i.e., in terms of Mori's original derivation, the generalized Langevin
equation for the linearly independent basis functions ${\bf b}(t)$ is given by
\begin{equation}
\frac{d{\bf g}(t)}{dt}={\bf M}\cdot{\bf g}(t)+{\cal N}(t)-\int_0^t{\bf K}(t-s)
\cdot{\bf g}(s)ds\;.
\end{equation}
The advantage of Mori’s projection operator is that, due to the linearity of
the projected low-dimensional functions, the derivation of the kernel {\bf K} 
is significantly simplified.

Extensive data were obtained by numerical simulation of fully-resolved discrete
Eulerian Navior-Stokes equations, given by
\begin{equation}
\frac{\partial v_i}{\partial t}+\frac{\partial v_iv_j}{\partial x_j}=-
\frac{\partial p}{\partial x_i}+\nu\frac{\partial^2 v_i}{\partial x_j^2}\;,
\end{equation}
where $\nu$ is the kinematic viscosity, and $p$ is the pressure that was
computed by solving the Poisson's equation for $p$. The data were used to 
extract the kernel and the noise term in Eq. (53) by computing a two-point 
correlation function and relating them to each other by an iterative process 
[174]. Figure 13 compares the Frobenius norm of the memory kernal (normalized 
by its corresponding Markov operator) for a set of observable in the original 
data with the results obtained with a Gaussian filter of various resolution, as
measured by filtering length $l_\Delta$. The Frobenius norm of an $m\times n$ 
matrix is defined as the square root of the sum of the squares of its elements.
As discussed by Tian {\it et al.} [171], the Frobenius norm of the memory 
kernel does not vanish with a finite time delay, but becomes two to three 
orders of magnitude smaller at a time delay around several Kolmogorov 
timescales (i.e., the smallest time scale in turbulent flow), hence indicating 
that using finite support in the memory integral can be a reasonable 
assumption, because the contributions from large time delays are generally 
negligible. Moreover, the effect of the filtering length scale $l_\Delta$ is 
significant. With larger $l_\Delta$ the temporal decay of the memory kernel
becomes slower, making the finite memory length longer, hence indicating a 
shift of dynamical contributions from the Markov term to memory integral. 

As mentioned above, even though the formalism was developed over 50 years ago, 
due to the intensive computations that are required for determining the kernel 
in the integro-differential equation that represents the generalized Lagevin 
equation, as well as the complexity of selecting the projection operator, only 
very recently have the applications of the method begun to emerge. They include
developing a reduced-order model for turbulence by Parish and Duraisamy 
[172,175] and Maeyama and Watanabe [176]. Li and Stinis [177] developed a 
reduced-order model for uncertainty quantification, while Stinis [178] 
presented a series of higher-order models for the Euler equation based on the 
Mori-Zwanzig formulation. The research field is finally emerging.

\begin{center}
{\bf E. Machine-Learning Approaches}
\end{center}

The approaches described so far in this section do not utlize machine-learning 
algorithms. There is a emerging class of data-driven approaches for discovering
the governing equations for complex phenomena that relies partly on such
algorithms. A good discussion of the issues that one must address when using 
machine learning to discover the governing equation for a dynamical system is 
given by Qin {\it et al.} [179].

One example of such approaches is the work of DiPeitro {\it et al.} [180], who
introduced a model for deriving the Hamiltonian of a dynamical system based
on data. Suppose that the Hamiltonian system is described by ${\bf q}=(q_1,
q_2,\cdots, q_n)$ and ${\bf p}=(p_1,p_2,\cdots,p_n)$, where {\bf q} and {\bf p}
represent, respectively, the position and momentum of ``object'' $i$ in the 
system. As usual, the evolution of the system is described by, $d{\bf p}/dt=-
\partial{\cal H}/\partial{\bf q}$ and $d{\bf q}/dt=\partial{\cal H}/\partial
{\bf p}$, where ${\cal H}$ is the Hamiltonian, or total energy, of the system, 
subject to the initial conditions ${\bf q}_0$ and ${\bf p}_0$. The time 
evolution is symplectomorphic, i.e., it conserves the volume form of the phase 
space and the symplectic 2-form wedge product $d{\bf p}\wedge{\bf x}$. DiPietro
{\it et al.} assumed that the Hamiltonian is separable, i.e., it can be written
as, ${\cal H}=E_p+E_k$, with $E_p$ and $E_k$ being the potential and kinetic 
energy.

Their approach, which they dubbed sparse symplectically integrated neural 
network, utilizes two neural networks, ${\cal N}_{E_p}$ and ${\cal N}_{E_k}$, 
which parametrize the potential and kinetic energies of the total Hamiltonian. 
Each network carries out a sparse regression (see above) within a search space 
specified by the user, which can include various functional forms, such as 
multivariate polynomials, trigonometric functions, and others, and computes the
terms of the function basis within the forward pass. The transformation must 
happen within the networks so as to enable the user to automatically compute 
gradients with respect to {\bf q} and {\bf p}. The basis terms are then passed 
through a single fully-connected layer, which learns the necessary terms of the
basis by making the trainable parameters to be the coefficients of each basis 
term, which are learned linearly with respect to each term in the basis. 
Depending on the spacified function space, one can modify the architecture of 
the networks. For example, one may employ an additional layers with bias if 
parameterizing using trigonometric functions 

For the purpose of training, as well as making predictions, the two networks 
are coupled with a symplectic integration scheme, which can of any order, 
depending on how much computing time one is willing or can afford to spend. 
DiPietro {\it et al.} [180] used a fourth-order integration scheme. Each time 
the gradients of the Hamiltonian (see above) are required, it is propagated 
through the networks, the necessary gradients are automatically computed, and 
are sent to the symplectic integrator. Since, depending on the size of the time
step, fourth-order symplectic integration often requires many iterative 
computations, one has frequently multiple passes through each network before 
the loss or cost function is computed. After the next state has been 
calculated, one computes the L1-norm between the predicted and the actual next 
state. L1-regularization is also incorporated so that only the essential
terms of the Hamiltonian survive. One can also achieve the same by using
thresholding that eliminates completely the non-essential terms. The loss
function is then defined and computed, and the optimization process for 
minimizing it is carried out.

Another approach is based on deep operator networks, DeepONets [181], which 
learn operators accurately and efficiently from a relatively small dataset in a
supervised data-driven manner. DeepONets consist of two sub-networks, one for 
encoding the input function at a fixed number of sensors $x_i,\;i=1,\cdots, m$,
which represents the branch net, and a second sub-network for encoding the 
locations for the output functions, the trunk net. One performs systematic 
simulations for identifying the PDE that governs the data. It has been 
demonstrated that DeepONet significantly reduces the generalization error, when
compared with the fully-connected neural networks.

Note that DeepONet is different from PIML algorithms described in Sec. IV, 
which are used to make predictions for various phenomena in complex media in
which the solution of a {\it known} PDE is modeled by a deep convolutional 
neural network whose parameters, together with other parameters of the model, 
are learned, but only constitutive relationships are discovered, since the 
fundamental underlying physics is established a priori. For example, Reyes 
{\it et al.} [182] used a PIML algorithm to discover viscosity models for two 
non-Newtonian systems, namely, polymer melts and suspensions of particles, in 
which they used only the data for the fluid velocity.

A hybrid method, DeepM\&Mnet, a composite supervised neural network, has also 
been proposed that combines DeepONets with the physics encoded by PIMLs, in 
order to obtain faster and more accurate solutions for complex problems. For 
example, Cai {\it et al.} [183] developed the approach to study 
electroconvection that results from coupling of a flow field with an electric 
field, as well as the concentration distributions of the cations and anions. In
their approach, given general inputs from the rest of the fields, one first 
pre-trains DeepONets that each field predicts independently. 

In another application, Mao {\it et al.} [184] used the same hybrid approach to
study high-speed flow past a normal shock. In this phenomenon the temperature 
of the fluid increases rapidly, triggerring chemical dissociation reactions
downstream. The spieces give rise to appreciable changes in the properties of 
the fluid. Hence, one has a coupled multiphysics multiscale dynamic phenomenon.
Carrying out standard numerical simulation of the phenomenon is extremely
difficult, whereas the hybrid DeepM\&Mnet can integrate seamlessly, given 
sparse measurements of the state variables in the simulation algorithm. 

\begin{center}
{\bf VIII. POSSIBLE FUTURE DIRECTIONS}
\end{center}

Our world is currently grappling with many highly difficult, but also 
tremendously important problems for which vast amnount of data are either 
already available, or are becoming so, but the physical laws, or more precisely
the equations that govern them, remain elusive. They include, but not limited 
to, understanding the neural basis of cognition and other biological systems, 
predicting large earthquakes, predicting the fate of contaminants in
groundwater aquifers, extracting and predicting coherent changes in the 
climate, understanding and predicting global soil salinization as severe 
drought afflicts large parts of the world, managing the spread of such emerging
diseases as COVID-19, controlling turbulence, and many more.

The goal of this Perspective was to describe recent progress in developing
theoretical and computational approaches that can not only meaningfully analyze
huge amounts of data, in a reasonable time, which contain information about the
properties of complex phenomena, such as those listed above, but also provide 
a framework for predicting their future behavior. As this Perspective has 
hopefully demonstrated, many approaches have been developed. But, although the 
``buzzword'' is that machine learning and artificial intelligence are going to 
solve many, if not all the problems listed above, that is not the case, at 
least not for the short and intermediate time scales. Artificial intelligence 
is not a panacea for all problems in science and engineering, and if it is not 
used the right way, it can create the misguided illusion that all the problems 
listed above and many more are going to be solved over the next 5-10 years, 
which is not the case and set science back.

At the same time, as this Perspective tried to make clear, there has been 
great progress in developing approaches that not only do not rely on machine 
learning, but have also provided new routes for dealing with big data that are 
becoming available all across science and engineering.  Thus, the question of 
which route to take is by itself a critical one. In some cases, such as climate
modeling that involves multiple widely disparate lenegh scales, as well as
extremely long times, the current computational power does not allow carrying
out numerical simulations over all the length and time scales. Therefore,
a combination of machine-learning algorithms and highly resolved, but 
affordable simulations, is perhaps the best route. Other cases represent 
``either'' or ``or'' system, whereby one can still deal with big data for them 
without resolrting to machine learning, or the training a neural network with 
suitable architecture may be the only hope. 

Even when it comes to the approaches that are currently available, while it is 
true that tremendous progress has been made in about a decade or so, many 
problems remain. Some are purely theoretical, while many are practical issues
involving the speed of the computations, the range of parameter space that can 
be accessed, etc. For example,

(i) although the machine learning-based approaches have enjoyed tremendus 
success, a rigorous theoretical foundation as to why they are successful, or 
when they may fail, is still lacking. Thus, one needs new theories, and 
perhasps new mathematics, in order to analyze the limitations, as well as 
capabilities of physica- and data-informed algorithms. 

(ii) When it comes to the Mori-Zwanzig approach, the question of how to
efficiently and accurately construct the kernel and other terms of the 
formulations is still very much open.

(iii) Discovering the governing equations from sparse identification of 
nonlinear dynamical systems still has many hurdles to overcome. One must, for
example, address [120] the issue of the correct choice of measurement 
coordinates and of sparsifying function basis for the dynamics. There is no 
simple solution to this problem [120] and, theefore, a coordinated effort to 
incorporate expert knowledge, feature extraction, and other advanced methods is
needed.

(iv) Since many of the methods that were described, including symbolic
regression, and machine learning-based algorithms, involve use of stochastic 
optimization algorithms, one important question is whether it is possible to 
have no or extremely small training loss, when an optimization method is used. 
Other errors that need to be rigorously analyzed include those involved in the 
approximate solution of the PDEs, as well as the question that is often asked, 
namely, does a smaller training error imply more accurate predictions? 

(v) Many multiphysics and multiscale complex phenomena occur in systems with 
complicated geometry that must be incorporated into the algorithms, which is 
not an easy task due to the required computation time. Although some efforts 
have been made to address such questions [185-188], much remains to be 
explored. 

(vi) Even when it is clear one needs a synthesis of two or more approaches, say
a combination of a machine learning algorithm and intensive numerical 
simulation, one needs to be equipped with, for example, optimization theory
and theory of PDEs. In addition, there is always tremendous need for yet
faster numerical simulation and analysis. The combination of such branches of 
science is openning up new research venues.

In addition, every new approach or algorithm requires benchmarks for checking 
its accuracy and efficiency. When dealing with huge amounts of data for complex
phenomena and systems, such benchmarks must provide a meaningful evaluation of 
the algorithms. Selecting such benchmarks is also not an easy task and requires
careful considerations, as does the task of selecting the way by which such 
data should be made publicly available, a way that is accessible to a larger 
number of potential users.

In terms of moving in the direction of much wider use of such algorithms, we
recall that one reason that platforms for conventional computations, such as 
OpenFOAM [189] for simulation of fluid flow and transport processes, and the 
FEniCS [190] that solves differential equations by finite-element method, are 
popular is that they are user-friendly. Thus, for example, in the area of 
applications of the physics- and date-informed algorithms or the symbolic 
regression methods, in order to make such approaches ``everyday tools'' of 
research and development, they must also be user-friendly, and provide tools of
visualization and tracking the variables as they evolve in space and time.

\begin{center}
{\bf ACKNOWLEDGMENTS}
\end{center}

Over the past several years, I have benefitted greatly from stimulating 
discussions and fruitful collaboration with many colleague who worked with me 
on some of the problems described in this Perspective. I am particularly 
grateful to Felipe de Barros, Jinwoo Im, Serveh Kamrava, Joachim Peinke, Reza 
Rahimi Tabar, Pejman Tahmasebi, and Sami Masri.  

\newpage

\newcounter{bean}
\begin{list}%
{[\arabic{bean}]}{\usecounter{bean}\setlength{\rightmargin}{\leftmargin}}

\item J.H. Seinfeld and S.N. Pandis, {\it Atmospheric Chemistry and Physics} 
(Wiley, New York, 1998).

\item D. Simpson, Long-period modelling of photochemical oxidants in Europe.
Model calculations for July 1985, Atmos. Environ. {\bf 26}, 1609 (1992).

\item A. Heidarinasab, B. Dabir, and M. Sahimi, Multiresolution wavelet-based 
simulation of transport and photochemical reactions in the atmosphere, Atmos. 
Environ. {\bf 38}, 6381 (2004).

\item P. Tahmasebi, S. Kamrava, T. Bai, and M. Sahimi, Machine learning in geo-
and environmental sciences: From small to large scale, Adv. Water Resour. {\bf 
142}, 103619 (2020).

\item G.E. Karniadakis, I.G. Kevrekidis, L. Lu, P. Perdikaris, S. Wang, and L. 
Yang, Physics-informed machine learning, Nature Rev. Phys. {\bf 3}, 422 (2021).

\item M. Reichstein, G. Camps-Valls, B. Stevens, M. Jung, J. Denzler, N. 
Carvalhais, and Mr Prabhat, Deep learning and process understanding for 
data-driven earth system science, Nature {\bf 566}, 195 (2019).

\item S. Kamrava, P. Tahmasebi, and M. Sahimi, Simulating fluid Flow in complex
porous materials: Integrating the governing equations with deep-layered 
machines, NPJ Comput. Mater. {\bf 7}, 127 (2021).

\item M. Alber, A.B. Tepole, W.R. Cannon, S. De, S. Dura-Bernal, K. Garikipati,
G. Karniadakis, W.W. Lytton, P. Perdikaris, L. Petzold, and E. Kuhl, 
Integrating machine learning and multiscale modeling — perspectives, 
challenges, and opportunities in the biological, biomedical, and behavioral 
sciences, NPJ Digit. Med. {\bf 2}, 1 (2019).

\item M. Sahimi, {\it Flow and Transport in Porous Media and Fractured Rock},
2nd. ed. (Wiley-VCH, Weinheim, 2011).

\item M. Sahimi and S.E. Tajer, Self-affine distributions of the bulk density, 
elastic moduli, and seismic wave velocities of rock, Phys. Rev. E {\bf 71}, 
046301 (2005).

\item Z. Zhang and J.C. Moore, {\it Mathematical and Physical Fundamentals of 
Climate Change} (Elsevier, Amsterdam, 2015), Chapter 9.

\item G. Cressman, An operational objective analysis system, Mon. Wea. Rev.
{\bf 87}, 367 (1959).

\item R.E. Kalman, A new approach to linear filtering and prediction problems,
Trans. ASME, J. Basic Engineeering {\bf 87}, 35 (1960).

\item G. Evensen, Using the extended Kalman filter with a multilayer
quasi-geostrophic ocean model, J. Geophys. Res. {\bf 97}, 17905 (1992).

\item G. Evensen, Sequential data assimilation with a nonlinear
quasi-geostrophic model using monte carlo methods to forecast error statistics,
J. Geophys. Res. {\bf 99}, 10143 (1994).

\item P.L. Houtekamer and H.L. Mitchell, Data assimilation using an ensemble 
Kalman filter technique, Mont. Wea. Rev. {\bf 126}, 796 (1998).

\item P.L. Houtekamer and H.L. Mitchell, Ensemble Kalman filtering, Quarterly 
J. Roy. Meteorol. Soc. {\bf 131}, 3269 (2005).

\item H. Li, S.J. Qin, T.T. Tsotsis, and M. Sahimi, Computer simulation of gas 
generation and transport in landfills. VI. Dynamic updating of the model using 
the ensemble Kalman filter, Chem. Eng. Sci. {\bf 74}, 69 (2012).

\item H. Li, T.T. Tsotsis, M. Sahimi, and S.J. Qin, Ensembles-based and
GA-based optimization for landfill gas production, AIChE J. {\bf 60}, 2063
(2014).

\item See, for example, C.M. Bishop, Neural networks and their applications, 
Rev. Sci. Instrum. {\bf 65}, 1803 (1994)

\item S. Torquato, {\it Random Heterogeneous Materials} (Springer, New York,
2002).

\item M. Sahimi, {\it Heterogeneous Materials}, Volumes I and II (Springer,
New York, 2003).

\item S. Kamrava, P. Tahmasebi, and M. Sahimi, Linking morphology of porous 
media to their macroscopic permeability by deep learning, Transp. Porous Media 
{\bf 131}, 427 (2020).

\item S. Kamrava, J. Im, F.P.J. de Barros, and M. Sahimi, Estimating dispersion
coefficient in flow through heterogeneous porous media by a deep convolutional 
neural network, Geophys. Res. Lett. {\bf 48}, e2021GL094443 (2021).

\item H. Wu, W.Z. Fang, Q. Kang, W.Q. Tao, and R. Qiao, Predicting effective 
diffusivity of porous media from images by deep learning, Sci. Rep. {\bf 9}, 
20387 (2019).

\item N. Alqahtani, F. Alzubaidi, R.T. Armstrong, P. Swietojanski, and P.
Mostaghimi, Machine learning for predicting properties of porous media from 2d 
X-ray images, J. Pet. Sci. Eng. {\bf 184}, 106514 (2020).

\item K.M. Graczyk and M. Matyka, Predicting porosity, permeability, and 
tortuosity of porous media from images by deep learning, Sci. Rep. {\bf 10}, 
21488 (2020).

\item L. Zhou, L. Shi, and Y. Zha, Seeing macro-dispersivity from hydraulic 
conductivity field with convolutional neural network, Adv. Water Resour. 
{\bf 138}, 103545 (2020).

\item S. Kamrava, P. Tahmasebi, and M. Sahimi, Enhancing images of shale 
formations by a hybrid stochastic and deep learning algorithm, Neural Networks 
{\bf 118}, 310 (2019).

\item S. Kamrava, P. Tahmasebi, and M. Sahimi, Physics- and image-based 
prediction of fluid flow and transport in complex porous membranes and 
materials by deep learning, J. Membr. Sci. {\bf 622}, 119050 (2021).

\item S. Ioffe and C. Szegedy, Batch normalization: Accelerating deep network 
training by reducing internal covariate shift, arXiv:1502.03167V3 (2015).

\item S. Kullback and R.A. Leibler, On information and sufficiency, Annal. 
Math. Statist. {\bf 22}, 79 (1951).

\item H. Andre\"a, N. Combaret, J. Dvorkin, E. Glatt, J. Han, M. Kabel, Y.
Keehm, F. Krzikalla, M. Lee, C. Madonna, M. Marsh, T. Mukerji, E.H. Saenger, R.
Sain, N. Saxena, S. Ricker, A. Wiegmann, and X. Zhan, Digital rock physics 
benchmarks—Part I: imaging and segmentation, Comput. Geosci. {\bf 50}, 25 
(2013).

\item G. Kissas, Y. Yang, E. Hwuang, W.R. Witschey, J.A. Detre, and P.
Perdikaris, Machine learning in cardiovascular flows modeling: predicting 
arterial blood pressure from non-invasive 4D flow MRI data using 
physicsinformed neural networks, Comput. Methods Appl. Mech. Eng. {\bf 358}, 
112623 (2020)

\item X. Glorot and Y. Bengio, Understanding the difficulty of training deep 
feedforward neural networks, in, {\it Proceedings of the thirteenth 
international conference on artificial intelligence and statistics}, AISTATS 
2010, Chia Laguna Resort, Sardinia, Italy (May 2010), 249-256.

\item Y. Zhu, N. Zabaras, P.S. Koutsourelakis, and P. Perdikaris, 
Physics-constrained deep learning for high-dimensional surrogate modeling and 
uncertainty quantification without labeled data, J. Comput. Phys. {\bf 394}, 56
(2019).

\item N. Geneva and N. Zabaras, Modeling the dynamics of PDE systems with 
physics-constrained deep auto-regressive networks, J. Comput. Phys. {\bf 403}, 
109056 (2020).

\item J.L. Wu, K. Kashinath, A. Albert, D. Chirila, Prabhat, and H. Xiao,
Enforcing statistical constraints in generative adversarial networks for 
modeling chaotic dynamical systems, J. Comput. Phys. {\bf 406}, 109209 (2020).

\item A. Kashefi, D. Rempe, and L.J. Guibas, A point-cloud deep learning 
framework for prediction of fluid flow fields on irregular geometries, Phys. 
Fluids {\bf 33}, 027104 (2021).

\item Y. LeCun and Y. Bengio, Convolutional networks for images, speech, and 
time series, in {\it The Handbook of Brain Theory and Neural Networks}, edited
by M.A. Arbib (MIT Press, Cambridge, (1995).

\item J. Winkens, J. Linmans, B.S. Veeling, T.S. Cohen, and M. Welling, 
Improved semantic segmentation for histopathology using rotation equivariant 
convolutional networks, in {\it Proceedings of the 1st Conference on Medical 
Imaging with Deep Learning (MIDL 2018)}, Amsterdam, The Netherlands (2018).

\item T. Cohen, M. Weiler, B. Kicanaoglu, and M. Welling, Gauge equivariant 
convolutional networks and the icosahedral CNN, in {\it Proceedings of the 36th
International Conference on Machine Learning}, Long Beach, California, {\bf 
97}, 1321 (2019).

\item H. Owhadi and G.R. Yoo, Kernel flows: from learning kernels from data 
into the abyss, J. Comput. Phys. {\bf 389}, 22 (2019).

\item M. Raissi, P. Perdikaris, and G.E. Karniadakis, Numerical Gaussian 
processes for time-dependent and nonlinear partial differential equations, SIAM
J. Sci. Comput. {\bf 40}, A172 (2018).

\item B. Hamzi and H. Owhadi, Learning dynamical systems from data: a simple 
cross-validation perspective, part I: parametric kernel flows, Physica D {\bf 
421}, 132817 (2021).

\item S. Wang, X. Yu, and P. Perdikaris, When and why PINNs fail to train: a 
neural tangent kernel perspective, J. Comput. Phys. {\bf 449}, 110768 (2022).

\item S. Wang, H. Wang, and P. Perdikaris, On the eigenvector bias of Fourier 
feature networks: from regression to solving multi-scale PDEs with 
physics-informed neural networks, Comput. Methods Appl. Mech. Eng. {\bf 384},
113938 (2021).

\item H. Owhadi, Do ideas have shape? Plato’s theory of forms as the continuous
limit of artificial neural networks, https://arxiv.org/abs/2008.03920 (2020).

\item J. Zhou, G. Cui, S. Hu, Z. Zhang, C. Yang, Z. Liu, L. Wang, C. Li, and
M. Sun, Graph neural networks: A review of methods and applications, AI Open
{\bf 1}, 57 (2020).

\item A. Mathews, M. Francisquez, J. Hughes, and D. Hatch, Uncovering edge 
plasma dynamics via deep learning from partial observations, Phys. Rev. E {\bf 
104}, 025205 (2021).

\item D. Pfau, J.S. Spencer, A.G. Matthews, and W.M.C. Foulkes, Ab initio 
solution of the many-electron Schr\"odinger equation with deep neural networks,
Phys. Rev. Res. {\bf 2}, 033429 (2020).

\item G.P. Pun, R. Batra, R. Ramprasad, and Y. Mishin, Physically informed 
artificial neural networks for atomistic modeling of materials, Nat. Commun. 
{\bf 10}, 2339 (2019).

\item D. Li, K. Xu, J.M. Harris, and E. Darve, Coupled time-lapse full-waveform
inversion for subsurface flow problems using intrusive automatic 
differentiation, Water Resour. Res. {\bf 56}, e2019WR027032 (2020).

\item W. Zhu, K. Xu, E. Darve, and G.C. Beroza, A general approach to seismic 
inversion with automatic differentiation, Computer Geosci. {\bf 151}, 104751
(2021).

\item J. Behler and M. Parrinello, Generalized neural-network representation of
high-dimensional potential-energy surfaces, Phys. Rev. Lett. {\bf 98}, 146401 
(2007).

\item L. Zhang, J. Han, H. Wang, R. Car, and E. Weinan, Deep potential 
molecular dynamics: a scalable model with the accuracy of quantum mechanics, 
Phys. Rev. Lett. {\bf 120}, 143001 (2018).

\item W. Jia, H. Wang, M. Chen, D. Lu, L. Lin, R. Car, E. Weinan, and L. Zhang,
Pushing the limit of molecular dynamics with ab initio accuracy to 100 million 
atoms with machine learning, arXiv:2005.00223 (2020).

\item A. Nakata, J.S. Baker, S.Y. Mujahed, J.T.L. Poulton, S. Arapan, J. Lin,
Z. Raza, S. Yadav, L. Truflandier, T. Miyazaki, and D.R. Bowler, Large scale 
and linear scaling DFT with the CONQUEST code, J. Chem. Phys. {\bf 152}, 164112
(2020).

\item R. Mantegna and H. E. Stanley, {\it An Introduction to Econophysics: 
Correlations and Complexities in Finance} (Cambridge University Press, New 
York, 2000).

\item P. Manshoor, S. Saberi, M. Sahimi, J. Peinke, A.F. Pacheco, and M.R. 
Rahimi Tabar, Turbulencelike behavior of seismic time series, Phys. Rev. Lett.
{\bf 102}, 014101 (2009).

\item P.Ch. Ivanov, L.A. Amaral, A.L. Goldberger, S. Havlin, M.G. Rosenblum, 
Z.R. Struzik, and H.E. Stanley, Multifractality in human heartbeat dynamics,
Nature (London) {\bf 399}, 461 (1999).

\item Y. Ashkenazy, P.Ch. Ivanov, S. Havlin, C.-K. Peng, A.L. Goldberger, and 
H.E. Stanley, magnitude and sign correlations in heartbeat fluctuations, Phys. 
Rev. Lett. {\bf 86}, 1900 (2001).

\item J.D. Hamilton, {\it Time Series Analysis } (Princeton University Press, 
Princeton, 1994).

\item H. Kantz and T. Schreiber, {\it  Nonlinear Time Series Analysis} 
(Cambridge University Press, London, 2003).

\item R.H. Stoffer and S. David {\it Time Series Analysis and Its Applications}
(Springer-Verlag, Berlin, 2006).

\item R. Friedrich and J. Peinke, Description of a turbulent cascade by a 
Fokker-Planck equation, Phys. Rev. Lett. {\bf 78}, 863 (1997).

\item J. Davoudi and M.R. Rahimi Tabar, Theoretical model for the Kramers-Moyal
description of turbulence cascades, Phys. Rev. Lett. {\bf 82}, 1680 (1999).

\item R. Friedlich, J. Peinke, M. Sahimi, and M.R. Rahimi Tabar, Approaching 
complexity by stochastic methods: From biological systems to turbulence, Phys. 
Rep. {\bf 506}, 87 (2011).

\item B.B. Mandelbrot and J.W. van Ness, Fractional Brownian motions, 
fractional noises, and applications, SIAM Rev. {\bf 10}, 422 (1968).

\item G.R. Jafari, S.M. Fazeli, F. Ghasemi, S.M. Vaez Allaei, M.R. Rahimi 
Tabar, A. Iraji zad, and G. Kavei, stochastic analysis and regeneration of 
rough surfaces, Phys. Rev. Lett. {\bf 91}, 226101 (2003).

\item F. Ghasemi, M. Sahimi, J. Peinke, and M.R. Rahimi Tabar, Analysis of 
non-stationary data for heart-rate fluctuations in terms of drift and diffusion
coefficients, J. Biol. Phys. {\bf 32}, 117 (2006).

\item H. Risken, {\it The Fokker-Planck Equation}, 2nd ed. (Springer, Berlin, 
1996).

\item F. Ghasemi, M. Sahimi, J. Peinke, R. Friedrich, G.R. Jafari, and M.R. 
Rahimi Tabar, Markov analysis and Kramers-Moyal expansion of nonstationary 
stochastic processes with an application to the fluctuations in the oil price,
Phys. Rev. E {\bf 75}, 060102(R) (2007).

\item J.P. Bouchaud and R. Cont, A Langevin approach to stock market 
fluctuations and crashes, Eur. Phys. J. B {\bf 6}, 543 (1998).

\item F. Ghasemi, J. Peinke, M. Sahimi, and M.R. Rahimi Tabar, Regeneration of 
stochastic processes: An inverse method, Euro. Phys. J. B {\bf 47}, 411 (2005).

\item R.F. Pawula, Approximation of the linear Boltzmann equation by the 
Fokker-Planck equation, Phys. Rev. {\bf 162}, 186 (1967).

\item M. Anvari, K. Lehnertz, M.R. Rahimi Tabar, and J. Peinke, Disentangling 
the stochastic behavior of complex time series, Sci. Rep. {\bf 6}, 35435 
(2016).

\item M.R. Rahimi Tabar, {\it Analysis and Data-Based Reconstruction of Complex
Nonlinear Dynamical Systems: Using the Methods of Stochastic Processes} 
(Springer, Bern, 2019).

\item M.T. Giraudo and L. Sacerdote, Jump-diffusion processes as models for 
neuronal activity, Biosystems {\bf 40}, 75 (1997).

\item  R. Sirovich, L. Sacerdote, and A.E.P. Villa, Cooperative behavior in a 
jump diffusion model for a simple network of spiking neurons, Math. Biosci. 
Eng. {\bf 11}, 385 (2014).

\item E. Daly and A. Porporato, Probabilistic dynamics of some jump-diffusion 
systems, Phys. Rev. E {\bf 73}, 026108 (2006).

\item R. Cont and P. Tankov, {\it Financial Modelling with Jump Processes} 
(Chapman \& Hall, Boca Raton, 2004).

\item J. Prusseit and K. Lehnertz, Stochastic qualifiers of epileptic brain 
dynamics, Phys. Rev. Lett. {\bf 98}, 138103 (2007).

\item K. Lehnertz, Epilepsy and nonlinear dynamics, J. Biol. Phys. {\bf 34}, 
253 (2008).

\item L.R. Gorj\~ao, J. Heysel, K. Lehnertz, and M.R. Rahimi Tabar, Analysis 
and data-driven reconstruction of bivariate jump-diffusion processes, Phys. 
Rev. E {\bf 100}, 062127 (2019)

\item F. Nikakhtar, L. Parkavosi, M.R. Rahimi Tabar, M. Sahimi, K. Lehnertz, 
and U. Feudel, Data-driven reconstruction of stochastic dynamical equations 
based on statistical moments, Phys. Rev. Lett. (to be published).

\item M.R. Rahimi Tabar, F. Nikakhtar, L. Parkavosi, A. Akhshi, K. Lehnertz, 
and U. Feudel, Revealing higher-order interactions in high-dimensional complex 
systems: A data-driven approach, to be published.

\item F. Hourdin, T. Mauritsen, A. Gettelman, J.-C. Golaz, V. Balaji, Q. Duan, 
D. Folini, D. Ji, D. Klocke, Y. Qian, F. Rauser, C. Rio, L. Tomassini, M.
Watanabe, and D. Williamson, The art and science of climate model tuning, Bull.
Am. Meteorol. Soc. {\bf 98}, 589 (2017)

\item S. Scher, Toward data-driven weather and climate forecasting: 
Approximating a simple general circulation model with deep learning, Geophys.
Res. Lett. {\bf 45}, 12 (2018).

\item P.D. Dueben and P. Bauer, Challenges and design choices for global 
weather and climate models based on machine learning, Geosci. Model 
Development {\bf 11}, 3999 (2018).

\item D.C. Park, A time series data prediction scheme using bilinear recurrent 
neural network, in {\it Proceedings of the 2010 International Conference on 
Information Science and Applications}, Seoul, South Korea (2010), p. 1, doi:
10.1109/ICISA.2010.5480383.

\item R. Fablet, S. Ouala, and C. Herzet, Bilinear residual neural network for 
the identification and forecasting of geophysical dynamics, in {\it Proceedings
of 26th European Signal Processing Conference (EUSIPCO)}, Rome, Italy (2018), 
p. 1477, doi: 10.23919/EUSIPCO.2018.8553492.

\item K. Pathak, B. Hunt, M. Girvan, Z. Lu, and E. Ott, Model-free prediction 
of large spatiotemporally chaotic systems from data: A reservoir computing 
approach, Phys. Rev. Lett. {\bf 120}, 024102 (2018).

\item P. Laloyaux, M. Bonavita, M. Dahoui, J. Farnan, S. Healy, E. H\'olm, and
S. Lang, Towards an unbiased stratospheric analysis, Q. J. Roy. Meteorol. Soc.
{\bf 146}, 2392 (2020).

\item S. Rasp, M.S. Pritchard, and P. Gentine, Deep learning to represent 
subgrid processes in climate models, Proc. Natl. Acad. Sci. USA {\bf 115}, 
9684 (2018).

\item W.D. Collins et al., The formulation and atmospheric simulation of the
community atmosphere model version 3 (CAM3), J. Clim. {\bf 19}, 2144 (2006).

\item J. Brajard, A. Carrassi, M. Bocquet, and L. Bertino, Combining data 
assimilation and machine learning to infer unresolved scale parametrisation,
Phil. Trans. Roy. Soc. London {\bf 379}, 20200086 (2021).

\item L. De Cruz, J. Demaeyer, and S. Vannitsem, The modular arbitrary-order 
ocean-atmosphere model: MAOOAM v1.0, Geosci. Model Development {\bf 9}, 2793 
(2016).

\item J. Bongard and H. Lipson, Automated reverse engineering of nonlinear 
dynamical systems, Proc. Natl. Acad. Sci. U.S.A. {\bf 104}, 9943 (2007).

\item M. Schmidt and H. Lipson, Distilling free-form natural laws from 
experimental data, Science {\bf 324}, 81 (2009).

\item J. Bongard, V. Zykov, and H. Lipson, Resilient machines through 
continuous self-modeling, Science {\bf 314}, 1118 (2066).

\item M. Sahimi and P. Tahmasebi, Reconstruction, optimization, and design of 
heterogeneous materials and media: Basic principles, computational algorithms, 
and applications, Phys. Rep. {\bf 939}, 1 (2021).

\item S. Kirkpatrick, C.D. Gelatt, Jr., and M.P. Vecchi, Optimization by 
simulated annealing, Science {\bf 220}, 671 (1983).

\item S. Katoch, S. Singh Chauhan, and V. Kumar, A review on genetic algorithm:
past, present, and future, Multimed. Tools. Appl. {\bf 80}, 8091 (2021). 

\item S.J. Russell and P. Norvig, {\it Artificial Intelligence: A Modern 
Approach}, 2nd ed. (Prentice Hall, Upper Saddle River, 2003) pp. 111–114.

\item J. Im, F.P.J. de Barros, S. Masri, M. Sahimi, and R.M. Ziff, Data-driven 
discovery of the governing equations for transport in heterogeneous media by 
stochastic optimization, Phys. Rev. E, to be published

\item Y. Gefen, A. Aharony, and S. Alexander, Anomalous diffusion on 
percolation clusters, Phys. Rev. Lett. {\bf 50}, 77 (1983).

\item D. Stauffer and A. Aharony, {\it Introduction to Perclation Theory}, 2nd.
ed. (Taylor and Francis, London, 1994).

\item M. Sahimi, {\it Applications of Percolation Theory}, 2nd. ed. (Springer, 
New York, 2023).

\item B. O'Shaughnessy and I. Procaccia, Analytical solutions for diffusion on
fractal objects, Phys. Rev. Lett. {\bf 54}, 455 (1985).

\item M. Giona and H.E. Roman, Fractional diffusion equation for transport 
phenomena in random media, Physica A {\bf 185}, 87 (1992).

\item R. Metzler, W.G. Gl\"ockle, and T.F. Nonnenmacher, Fractional model 
equation for anomalous diffusion, Physica A {\bf 211}, 13 (1994).

\item Y. He, S. Burov, R. Metzler, and E. Barkai, Random time-scale invariant 
diffusion and transport coefficients, Phys. Rev. Lett. {\bf 101}, 058101 
(2008).

\item A. Bunde and J. Dr\"ager, Localization in disordered structures: 
Breakdown of the self-averaging hypothesis, Phys. Rev. E {\bf 52}, 53 (1995).

\item A. Pacheco-Pozo and I.M. Sokolov, Universal fluctuations and ergodicity
of generalized diffusivity on critical percolation clusters, J. Phys. A {\bf 
55}, 345001 (2022).

\item K. Garbrecht, M. Aguilo, A. Sanderson, A. Rollett, R.M. Kirby, and J.
Hochhalter, Interpretable machine learning for texture-dependent constitutive 
models with automatic code generation for topological optimization, Integr. 
Mater. Manuf. Innov. {\bf 10}, 373 (2021)

\item P.J. Schmid, Dynamic mode decomposition of numerical and experimental 
data, J. Fluid Mech. {\bf 656}, 5 (2010).

\item B.O. Koopman, Hamiltonian systems and transformation in Hilbert space,
Proc. Natl. Acad. Sci. USA {\bf 17}, 315 (1931).

\item M.R. Jovanovi\'c, P.J. Schmid, and J.W. Nichols, Sparsity-promoting 
dynamic mode decomposition, Phys. Fluids {\bf 26}, 024103 (2014).

\item S.L. Brunton, J.L. Proctor, and J.N. Kutz, Discovering governing 
equations from data by sparse identification of nonlinear dynamical systems,
Proc. Natl. Acad. Sci. U.S.A. {\bf 113}, 3932 (2016).

\item F. Takens, Detecting strange attractors in turbulence, Lect. Notes. Math.
{\bf 898}, 366 (1981).

\item H. Ye, R.J. Beamish, S.M. Glaser, S.C. H. Grant, C.-H. Hsieh, L.J. 
Richards, J.T. Schnute, and G. Sugihara, Equation-free mechanistic ecosystem 
forecasting using empirical dynamic modeling, Proc. Natl. Acad. Sci. U.S.A. 
{\bf 112}, E1569 (2015).

\item G. Berkooz, P. Holmes, and J.L. Lumley, The proper orthogonal 
decomposition in the analysis of turbulent flows, Annu. Rev. Fluid Mech {\bf 25
}, 539 (1993).

\item D. Ruelle and F. Takens, On the nature of turbulence, Commun. Math. Phys.
{\bf 20}, 167 (1971).

\item C.P. Jackson, A finite-element study of the onset of vortex shedding in 
flow past variously shaped bodies, J. Fluid Mech. {\bf 182}, 23 (1987).

\item Z. Zebib, Stability of viscous flow past a circular cylinder, J. Eng.
Math. {\bf 21}, 155 (1987).

\item B.R. Noack, K. Afanasiev, M. Morzynsk, D. Tadmor, and F. Thiele, A 
hierarchy of low-dimensional models for the transient and post-transient 
cylinder wake, J. Fluid Mech. {\bf 497}, 335 (2003).

\item T. Colonius and K. Taira, A fast immersed boundary method using a 
nullspace approach and multi-domain far-field boundary conditions, Comput.
Methods Appl. Mech. Eng. {\bf 197}, 2131 (2008).

\item See, https://github.com/dynamicslab/pysindy.

\item B.M. de Silva, K. Champion, M. Quade, J.-C. Loiseau, J.N. Kutz, and S.L.
Brunton, PySINDy: a Python package for the sparse identification of nonlinear 
dynamics from data, J. Open Source Software {\bf 5}, 2104 (2020).

\item K. Champion, B. Lusch, J.N. Kutz, and S.L. Brunton, Data-driven discovery
of coordinates and governing equations, Proc. Natl. Acad. Sci. U.S.A. {\bf 
116}, 22445 (2019).

\item C. Rackauckas, Y. Ma, J. Martensen, C. Warner, K. Zubov, R. Supekar, D.
Skinner, A. Ramadhan, and A. Edelman, Universal differential equations for 
scientific machine learning, arXiv:2001.04385 (2020).

\item M. Gelbrecht, N. Boers, and J. Kurths, Neural partial differential 
equations for chaotic systems, New J. Phys. {\bf 23}, 043005 (2021).

\item M. Kalia, S.L. Brunton, H.G.E. Meijer, C. Brune, and J.N. Kutz, Learning 
normal form autoencoders for data-driven discovery of universal, 
parameter-dependent governing equations, arXiv:2106.05102 (2021).

\item M. Sorokina, S. Sygletos, and S. Turitsyn, Sparse identification for 
nonlinear optical communication systems: SINO method, Optics Express {\bf 24},
30433 (2016).

\item M. Dam, M. Brøns, J.J. Rasmussen, V. Naulin, and J.S. Hesthaven, Sparse 
identification of a predator-prey system from simulation data of a convection 
model, Phys. Plasmas {\bf 24}, 022310 (2017).

\item J.-C. Loiseau and S.L. Brunton, Constrained sparse Galerkin regression,
J. Fluid Mech. {\bf 838}, 42 (2018).

\item J.-C. Loiseau, B.R. Noack, and S.L. Brunton. Sparse reduced-order 
modeling: sensor-based dynamics to full-state estimation, J. Fluid Mech. 
{\bf 844}, 459 (2018).

\item L. Boninsegna, F. N\"uske, and C. Clementi, Sparse learning of stochastic
dynamical equations, J. Chem. Phys. {\bf 148}, 241723 (2018).

\item P. Gel$\beta$, S. Klus, J. Eisert, and C. Sch\"utte, Multidimensional 
approximation of nonlinear dynamical systems, J. Comput. Nonlinear Dyn. {\bf 
14}, 061006 (2019).

\item S. Thaler, L. Paehler, and N.A. Adams. Sparse identification of 
truncation errors, J. Comput. Phys. {\bf 397}, 108851 (2019).

\item K. Kaheman, E. Kaiser, B. Strom, J.N. Kutz, and S.L. Brunton, Learning 
discrepancy models from experimental data, arXiv:1909.08574v1 (2019).

\item J.-C. Loiseau, Data-driven modeling of the chaotic thermal convection in 
an annular thermosyphon, Theor. Comput. Fluid Dyn. {\bf 34}, 339 (2020).

\item S. Beetham and J. Capecelatro, Formulating turbulence closures using 
sparse regression with embedded form invariance, Phys. Rev. Fluid. {\bf 5},
084611 (2020).

\item M. Schmelzer, R.P. Dwight, and P. Cinnella, Discovery of algebraic 
Reynolds-stress models using sparse symbolic regression, Flow, Turbulence and 
Combustion {\bf 104}, 579 (2020).

\item B.M. de Silva, D.M. Higdon, S.L. Brunton, and J.N. Kutz, Discovery of 
physics from data: universal laws and discrepancies, Front. Artif. Intell. 
{\bf 3}, 25 (2020).

\item J.J. Bramburger and J.N. Kutz, Poincar\'e maps for multiscale physics 
discovery and nonlinear Floquet theory, Physica D {\bf 408}, 132479 (2020).

\item J.L. Callaham, J.-C. Loiseau, G. Rigas and S.L. Brunton, Nonlinear 
stochastic modelling with Langevin regression, Proc. Roy. Soc. Lond. A {\bf 
477}, 20210092 (2021).

\item J.J. Bramburger, J.N. Kutz, and S.L. Brunton, Data-driven stabilization 
of periodic orbits, IEEE Access {\bf 9}, 43504 (2021).

\item D.E. Shea, S.L. Brunton, and J.N. Kutz, SINDy-BVP: Sparse identification 
of nonlinear dynamics for boundary value problems, Phys. Rev. Res. {\bf 3},
023255 (2021).

\item S. Beetham, R.O. Fox, and J. Capecelatro, Sparse identification of 
multiphase turbulence closures for coupled fluid–particle flows, J. Fluid Mech.
{\bf 914}, A11 (2021).

\item A.A. Kaptanoglu, K.D. Morgan, C.J. Hansen, and S.L. Brunton, 
Physics-constrained, low-dimensional models for MHD: First-principles and 
data-driven approaches, Phys. Rev. E {\bf 104}, 015206 (2021).

\item N. Deng, B.R. Noack, M. Morzy\'nski, and L.R. Pastur, Galerkin force 
model for transient and post-transient dynamics of the fluidic pinball, J. 
Fluid Mech. {\bf 918}, A4 (2021).

\item Y. Guan, S.L. Brunton, and I. Novosselov, Sparse nonlinear models of 
chaotic electroconvection, Roy. Soc. Open Sci. {\bf 8}, 202367 (2021).

\item J.L. Callaham, S.L. Brunton, and J.-C. Loiseau, On the role of nonlinear 
correlations in reduced-order modeling, J. Fluid Mech. {\bf 938}, A1 (2022).

\item J. L. Callaham, G. Rigas, J.-C. Loiseau, and S.L. Brunton, An empirical 
mean-field model of symmetry-breaking in a turbulent wake, Sci. Adv. {\bf 8},
eabm4786 (2022).

\item H. Schaeffer and S.G. McCalla, Sparse model selection via integral terms,
Phys. Rev. E {\bf 96}, 023302 (2017).

\item D.A. Messenger and D.M. Bortz, Weak SINDy for partial differential 
equations, J. Comput. Phys. {\bf 443}, 110525 (2021).

\item D.R. Gurevich, P.A. Reinbold, and R.O. Grigoriev, Robust and optimal 
sparse regression for nonlinear PDE models, Chaos {\bf 29}, 103113 (2019).

\item P.A. Reinbold, D.R. Gurevich, and R.O. Grigoriev, Using noisy or 
incomplete data to discover models of spatiotemporal dynamics, Phys. Rev. E 
{\bf 101}, 010202 (2020).

\item P.K.A. Reinbold, L.M. Kageorge, M.F. Schatz, and R.O. Grigoriev, Robust 
learning from noisy, incomplete, high-dimensional experimental data via 
physically constrained symbolic regression, Nat. Commun. {\bf 12}, 3219 (2021).

\item E.P. Alves and F. Fiuza, Data-driven discovery of reduced plasma physics 
models from fully-kinetic simulations, Phys. Rev. Res. {\bf 4}, 033192 (2022).

\item H. Mori, Transport, collective motion, and Brownian motion, Prog. Theor.
Phys. {\bf 33}, 423 (1965).

\item R. Zwanzig, Nonlinear generalized Langevin equations, J. Stat. Phys. 
{\bf 9}, 215 (1973).

\item G.F. Mazenko, {\it Nonequilibrium Statistical Mechanics} (Wiley-VCH,
Weinheim, 2006).

\item D.J. Evans and G. Morriss, {\it Statistical Mechanics of Nonequilibrium 
Liquids} (Cambridge University Press, Cambridge, 2008).

\item C. Hi\'jon, P. Esp\~anol, E. Vanden-Eijnden, and R. Delgado-Buscalioni,
Mori-Zwanzig formalism as a practical computational tool, Faraday Discuss. 
{\bf 144}, 301 (2010).

\item A. J. Chorin, O. H. Hald, and R. Kupferman, Optimal prediction and the 
Mori-Zwanzig representation of irreversible processes, Proc. Natl. Acad. Sci. 
USA {\bf 97}, 2968 (2000).

\item S.K.J. Falkena, C. Quinn, J. Sieber, J. Frank, and H.A. Dijkstra, 
Derivation of delay equation climate models using the Mori-Zwanzig formalism,
Proc. R. Soc. A {\bf 475}, 20190075 (2019).

\item A. Gouasmi, E.J. Parish, and K. Duraisamy, A priori estimation of memory 
effects in reduced-order models of nonlinear systems using the Mori-Zwanzig 
formalism, Proc. R. Soc. LOnd. A {\bf 473}, 20170385 (2017).

\item Y. Tian, Y.T. Lin, M. Anghel, and D. Livescu, Data-driven learning of 
Mori-Zwanzig operators for isotropic turbulence, Phys. Fluids {\bf 33}, 125118 
(2021).

\item E.J. Parish and K. Duraisamy, Non-Markovian closure models for large eddy
simulations using the Mori-Zwanzig formalism, Phys. Rev. Fluids {\bf 2}, 014604
(2017).

\item W. Chu and X. Li, The Mori-Zwanzig formalism for the derivation of a
fluctuating heat conduction model from molecular dynamics, Commun. Math. Sci.
{\bf 17}, 539 (2019).

\item Y.T. Lin, Y. Tian, D. Livescu, and M. Anghel, Data-driven learning for 
the Mori-Zwanzig formalism: A generalization of the Koopman learning framework,
SIAM J. Appl. Dyn. Syst. {\bf 20}, 101137/21M1401759 (2021).

\item E.J. Parish and K. Duraisamy, A dynamic subgrid scale model for large 
eddy simulations based on the Mori-Zwanzig formalism, J. Comput. Phys. {\bf 
349}, 154 (2017).

\item S. Maeyama and T.-H. Watanabe, Extracting and modeling the effects of
small-scale fluctuations on large-scale fluctuations by Mori-Zwanzig projection
operator method, J. Phys. Soc. Jpn. {\bf 89}, 024401 (2020).

\item J. Li and P. Stinis, Mori-Zwanzig reduced models for uncertainty 
quantification, J. Comput. Dyn. {\bf 6}, 39 (2019).

\item P. Stinis, Higher order Mori-Zwanzig models for the Euler equations,
Multiscale Modeling Simul. {\bf 6}, 10.1137/06066504X (2007).

\item T. Qin, K. Wu, and D. Xiu, Data driven governing equations approximation
using deep neural networks, J. Comput. Phys. {\bf 395}, 620 (2019).

\item D.M. DiPietro, S. Xiong, and B. Zhu, Sparse symplectically integrated 
neural networks, in {\it Advances in Neural Information Processing Systems 33}
(NeurIPS 2020), edited by H. Larochelle, M. Ranzato, R. Hadsell, M.F. Balcan,
and H. Lin, 34th Conference on Neural Information Processing Systems, 
Vancouver, Canada (2020).

\item L. Lu, P. Jin, and G.E. Karniadakis, Learning nonlinear operators via 
DeepONet based on the universal approximation theorem of operators, Nat. 
Machine Intel. {\bf 3}, 218 (2021).

\item B. Reyes, A.A. Howard, P. Perdikaris, and A.M. Tartakovsky, Learning 
unknown physics of non-Newtonian fluids, Phys. Rev. Fluids {\bf 6}, 073301 
(2021).

\item S. Cai, Z. Wang, L. Lu, T.A. Zaki, and G.E. Karniadakis, DeepM\&Mnet: 
inferring the electroconvection multiphysics fields based on operator 
approximation by neural networks, J. Comput. Phys. {\bf 436}, 110296 (2020).

\item Z. Mao, L. Lu, O. Marxen, T.A. Zaki, and G.E. Karniadakis, DeepM\&Mnet 
for hypersonics: Predicting the coupled flow and finite-rate chemistry behind 
a normal shock using neural-network approximation of operators, J. Comput.
Phys. {\bf 447}, 110698 (2021)

\item Y. Shin, J. Darbon, and G.E. Karniadakis, On the convergence of physics 
informed neural networks for linear second-order elliptic and parabolic type 
PDEs, Commun. Comput. Phys. {\bf 28}, 2042 (2020).

\item S. Mishra and R. Molinaro, Estimates on the generalization error of 
physics informed neural networks (PINNs) for approximating PDEs, IMA J. Numer.
Analysis {\bf 42}, 981 (2022).

\item S. Mishra and R. Molinaro, Estimates on the generalization error of 
physics informed neural networks (PINNs) for approximating PDEs II: a class 
of inverse problems, arXiv:2007.01138v2 (2021)

\item Y. Shin, Z. Zhang, and G.E. Karniadakis, Error estimates of residual 
minimization using neural networks for linear PDEs, arXiv:2010.08019 (2020).

\item H. Jasak, A. Jemcov, and Z. Tukovi\'c, OpenFOAM: A C$^{++}$ library for 
complex physics simulations, International Workshop on Coupled Methods in 
Numerical Dynamics IUC, Dubrovnik, Croatia (September 2007).

\item M. Aln{\ae}s, J. Blechta, J. Hake, A. Johansson, B. Kehlet, A. Logg, C.
Richardson, J. Ring, M.E. Rohnes, and G.N. Wells, The FEniCS project version 
1.5, Arch. Numer. Softw. {\bf 3}, 9 (2015).

\end{list}%


\newpage

\begin{figure}[htp]
\centering
\includegraphics[width=\textwidth]{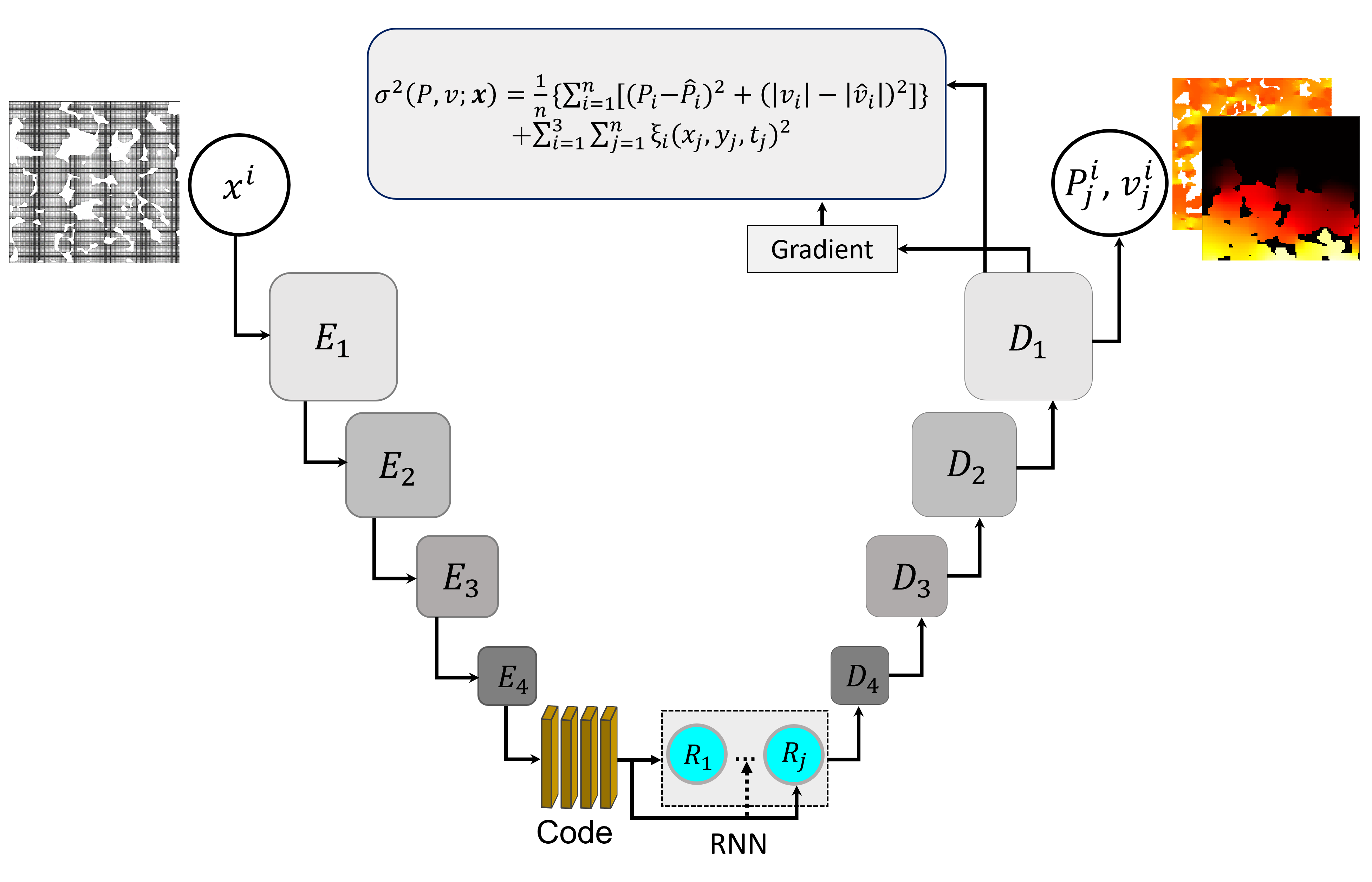}
\caption{The architecture of the PIRED network, with $E_i$ and $D_i$ indicating
the encoder and decoder blocks; $\sigma^2$ being the cost function, $x^i$ the 
input, and the pressure $P^j$ and fluid velocity $|v|^j$ are the output [7].}
\end{figure}

\newpage

\begin{figure}[htp]
\begin{center}
\includegraphics[width=\textwidth]{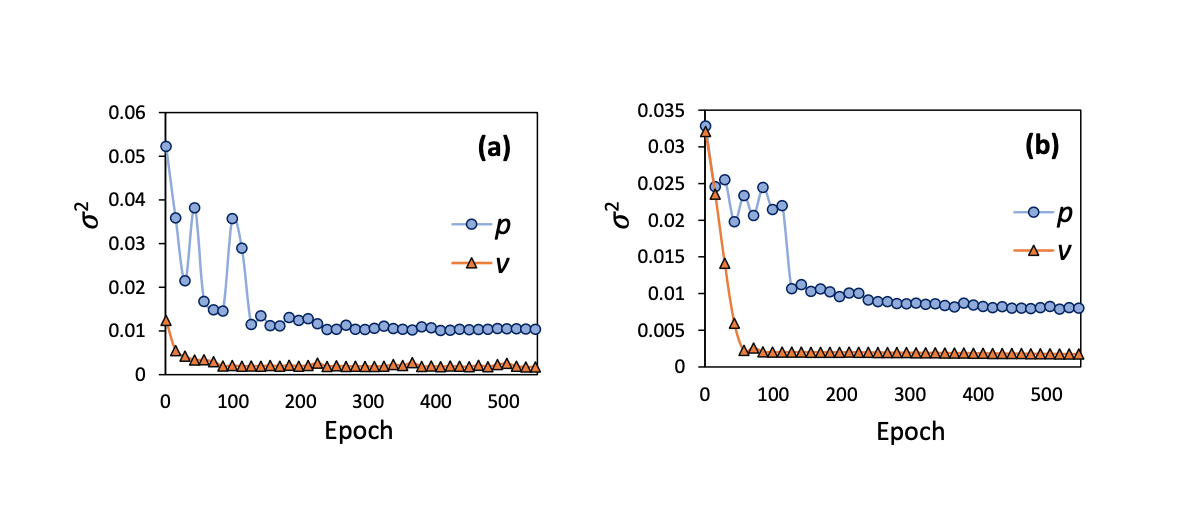}
\caption{Computational efficiency and accuracy of the PIRED. Comparison of the
cost $\sigma^2$ for (a) training and (b) test data of pressure $p$ and fluid 
velocity $v$.}
\end{center}
\end{figure}

\newpage

\begin{figure}[htp]
\begin{center}
\includegraphics[width=\textwidth]{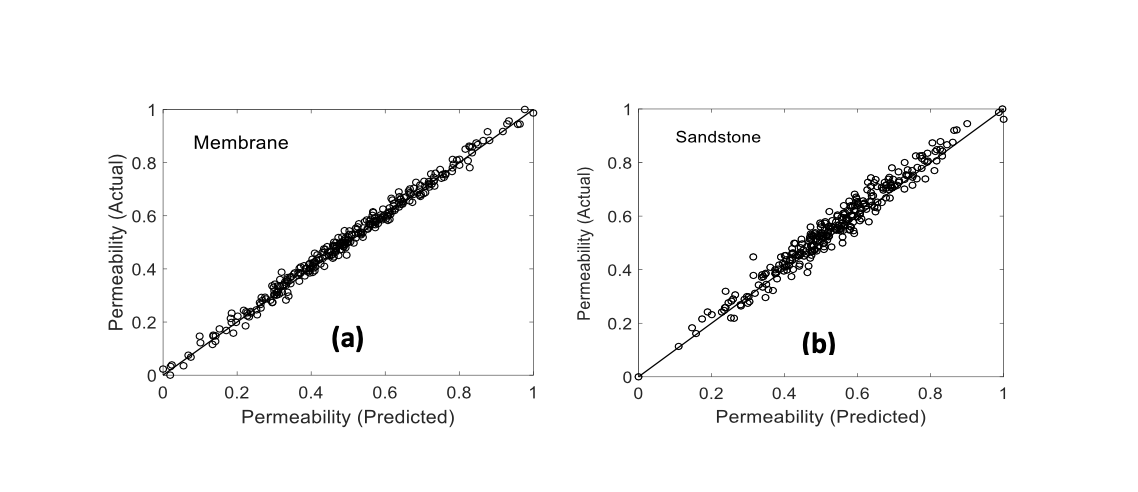}
\caption{Comparison of the actual permeabilities $K$ and the predictions by the
PIRED network for (a) 300 2D images of the membrane, and (b) for 100 images of 
the Fontainebleau sandstone. $K$ is normalized according to $(K-K_{\rm min})/
(K_{\rm max}-K_{\rm min})$ [7].}
\end{center}
\end{figure}

\newpage

\begin{figure}[htp]
\begin{center}
\includegraphics[width=\textwidth]{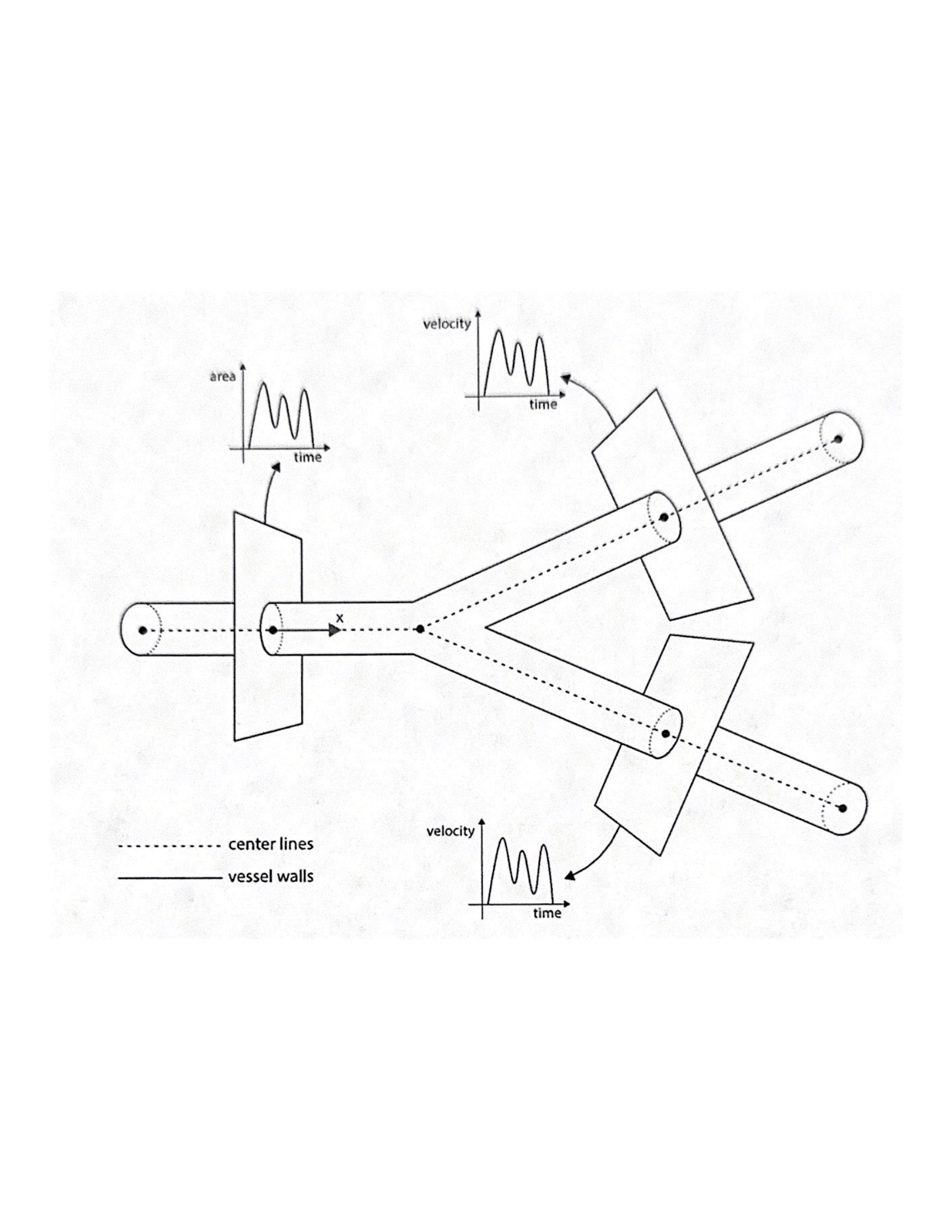}
\caption{Schematic representation of a $Y-$shaped bifurcating arterial system
and its 1D center-lines used in the reduced model [25].}
\end{center}
\end{figure}

\newpage

\begin{figure}[htp]
\begin{center}
\includegraphics[width=\textwidth]{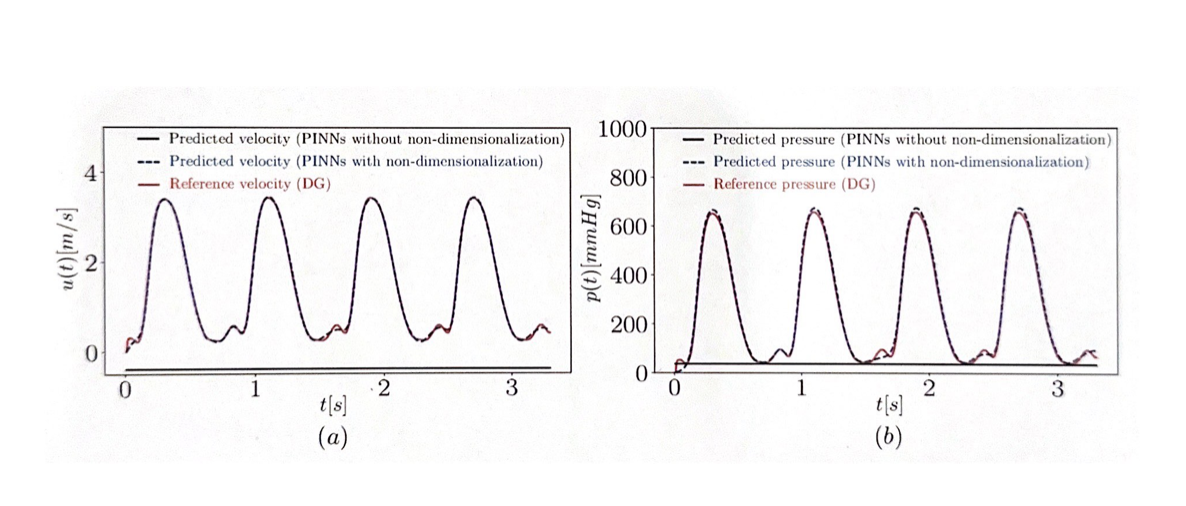}
\caption{Flow through a $Y-$shaped bifurcation. (a) Comparison of the computed
velocity, $u(t)=v_x(t)$ wave, obtained by the discontinuous Galerkin (DG, red) 
method with those predicted by the PINN with (blue) and without 
non-dimensionalization (black) at the middle point of channel 1 [the left 
channel in Fig. 4]. (b) Same as in (a), but for the pressure wave [25].}
\end{center}
\end{figure}

\newpage
\newpage
\newpage

\begin{figure}[htp]
\begin{center}
\includegraphics[width=\textwidth]{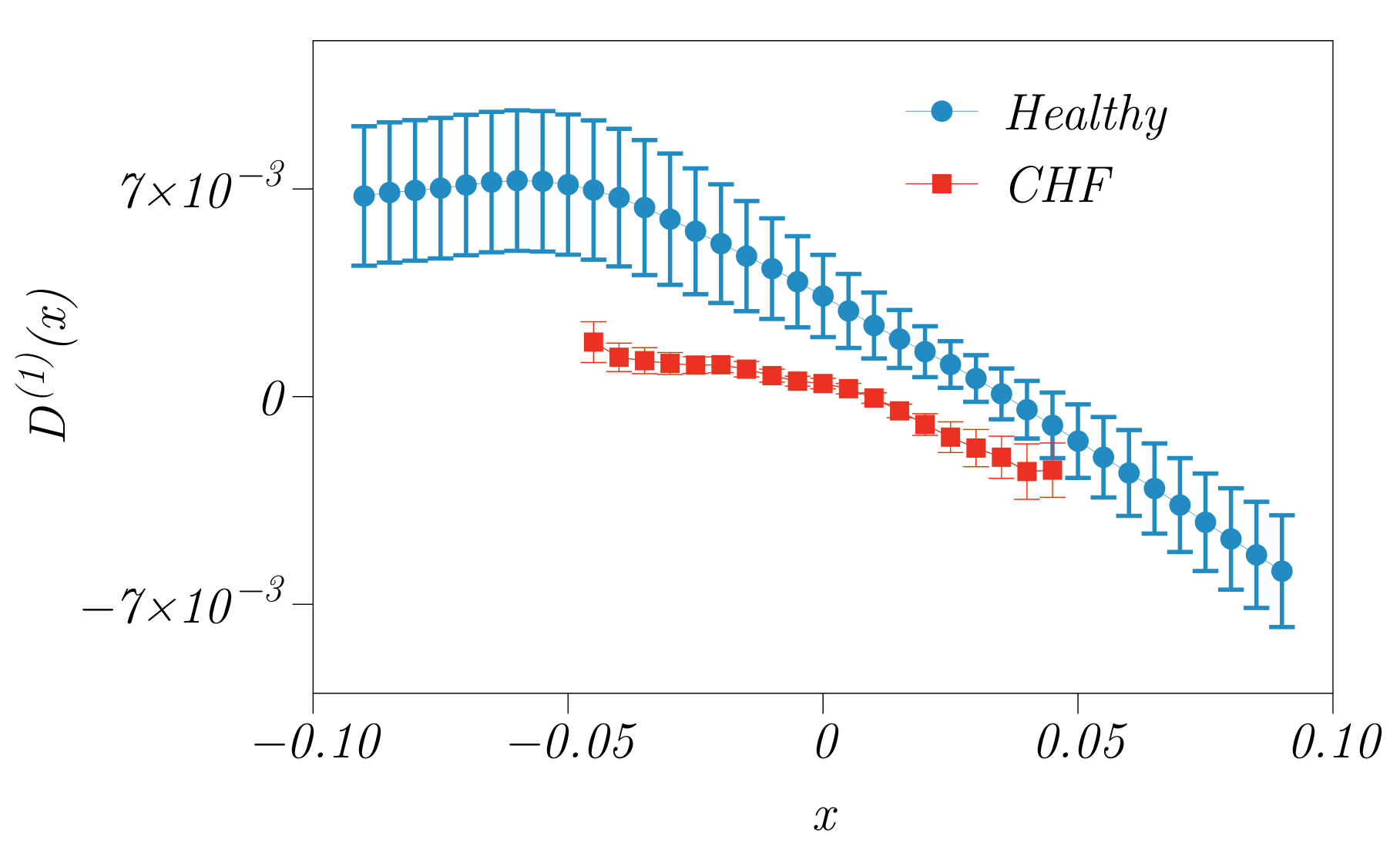}
\caption{The drift coefficient $D^{(1)}(x)$ for two classes of patients, the
healthy ones, and those with congestive heart failure (CHF).}
\end{center}
\end{figure}

\newpage

\begin{figure}[htp]
\begin{center}
\includegraphics[width=\textwidth]{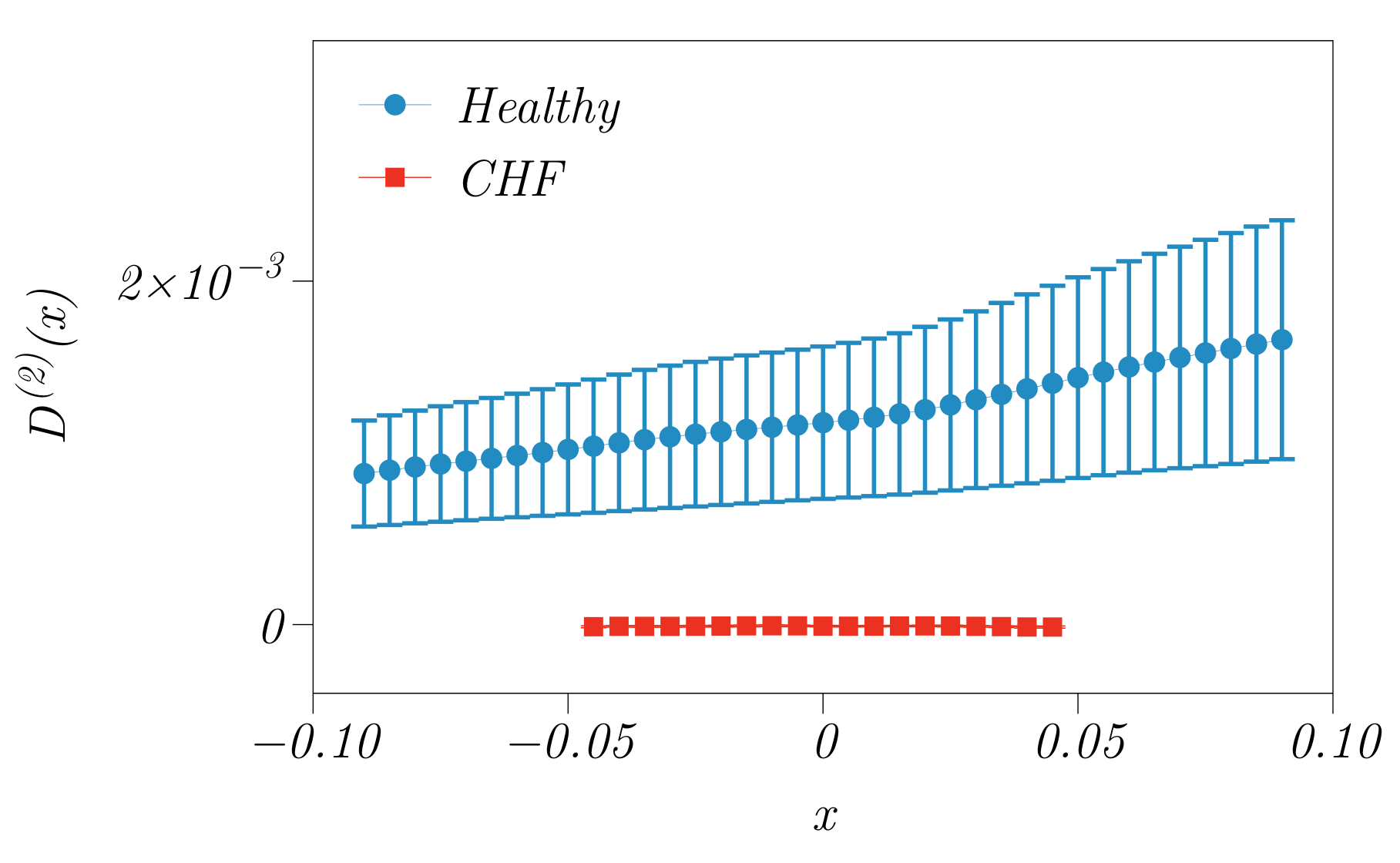}
\caption{The diffusion coefficient $D^{(2)}(x)$ for two classes of patients, 
the healthy ones, and those with congestive heart failure (CHF).}
\end{center}
\end{figure}

\newpage

\begin{figure}[htp]
\begin{center}
\includegraphics[width=\textwidth]{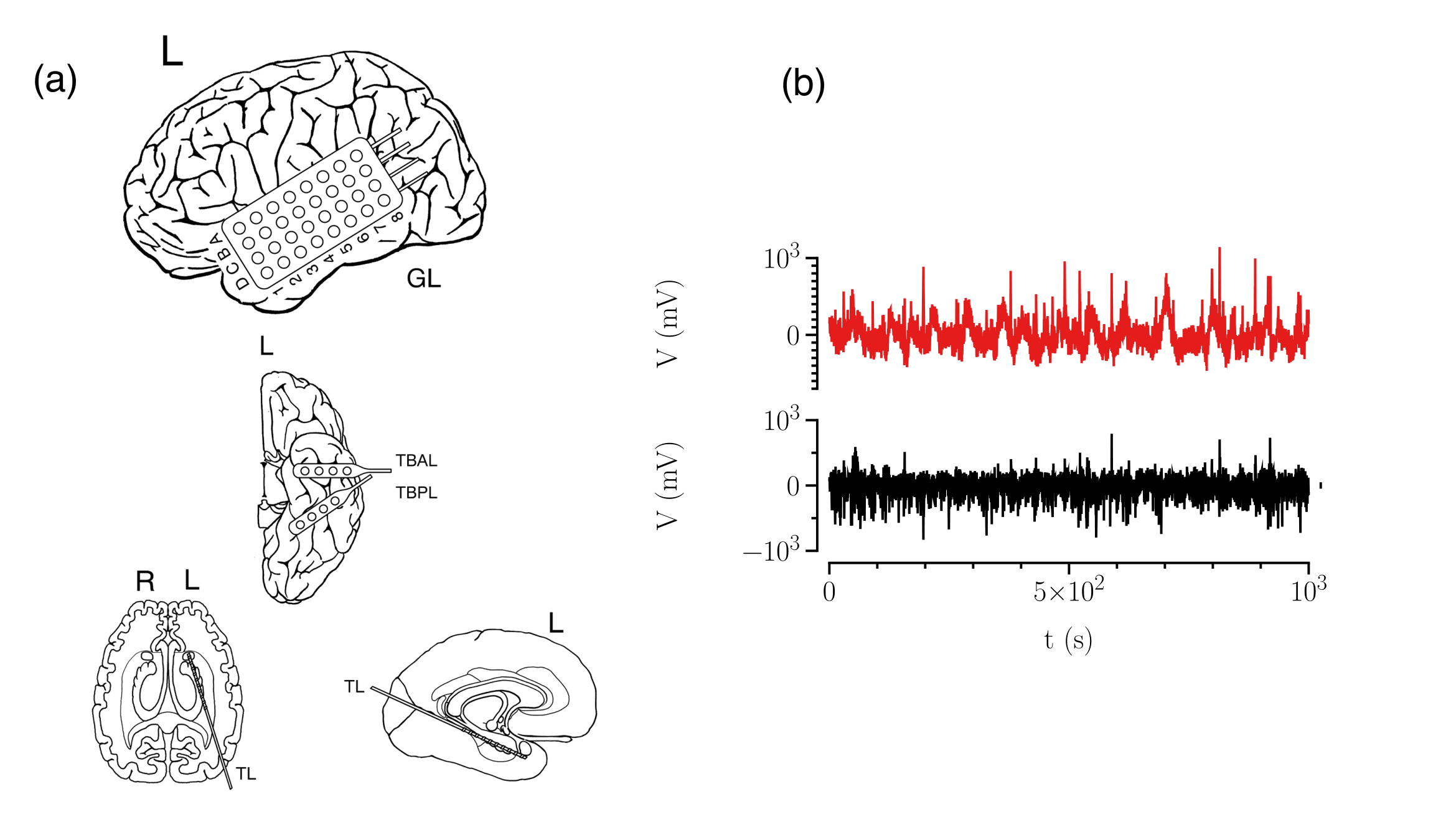}
\caption{(a) Implantation scheme of intracranial electrodes in a patient with 
seizures originating in the left mesial temporal lobe: temporal-lateral grid 
electrode ($8\time 4$ contacts, denoted by GL), two temporal-basal strip 
electrodes (4 contacts each, denoted by TB), and a hippocampal depth electrode 
(10 contacts, denoted by TL). The most anterior contact (TL1) is located 
ventral to the amygdala and the most posterior contact (TL10) is located within
the hippocampus. The latter electrode samples the epileptic focus. (b) Segments
of the iEEG time series recorded during the seizure-free interval from within 
the epileptic focus (red; contact TL4) and from a distant brain region 
(black; contact GLC6) [77].}
\end{center}
\end{figure}

\newpage

\begin{figure}[htp]
\begin{center}
\includegraphics[width=\textwidth]{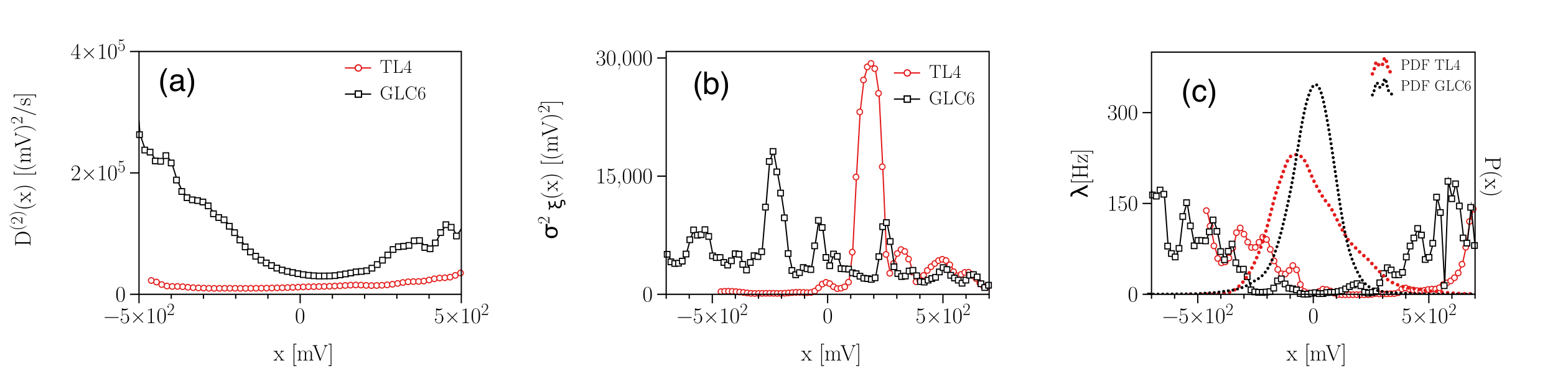}
\caption{Sample results computed based on the data for an epilepsy patient with
an epileptic focus in the left mesial temporal lobe, showing, (a) the diffusion
coefficients; (b) jump amplitudes, and (c) jump rates, together with the 
respective probability distribution functions estimated from normalized iEEG 
time series, recorded during the seizure-free interval from within the 
epileptic focus (red, contact TL4 in Fig. 8) and the data from a distant brain 
region (black, contact GLC6, as shown in Fig. 8) [77].}
\end{center}
\end{figure}

\pagebreak
\begin{figure}[htp]
\begin{center}
\includegraphics[width=\textwidth]{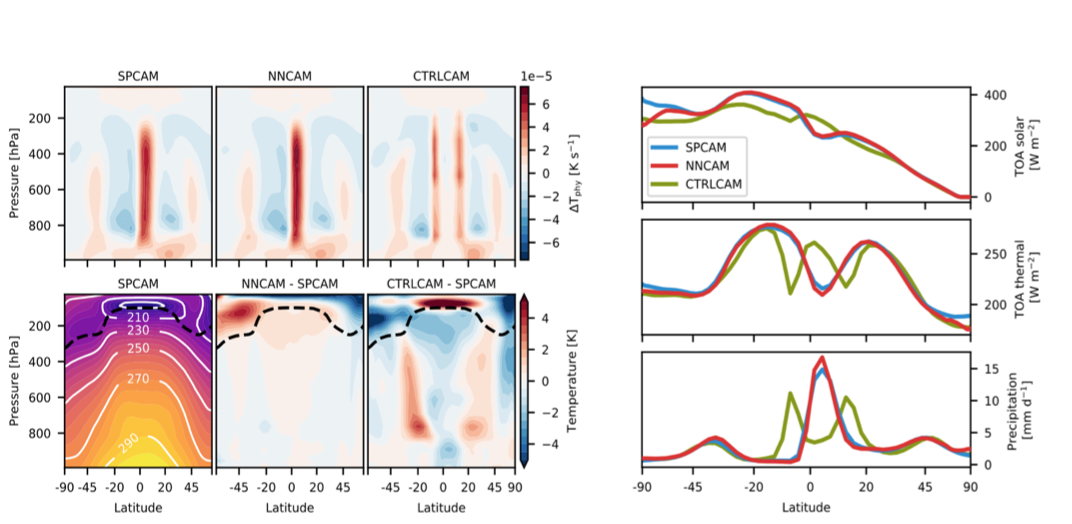}
\caption{Longitudinal and five-year temporal averages. (A) Mean convective and 
radiative subgrid heating rates $\Delta T_{\rm phy}$. (B) Mean temperature $T$ 
of SPCAM and biases of NNCAM and CTRLCAM relative to SPCAM. The dashed black 
line denotes the approximate position of the tropopause. (C) Mean short- 
(solar) and longwave (thermal) net fluxes at the top of the atmosphere and 
precipitation. The latitude axis is area weighted [95].}
\end{center}
\end{figure}

\newpage

\begin{figure}[htp]
\begin{center}
\includegraphics[width=\textwidth]{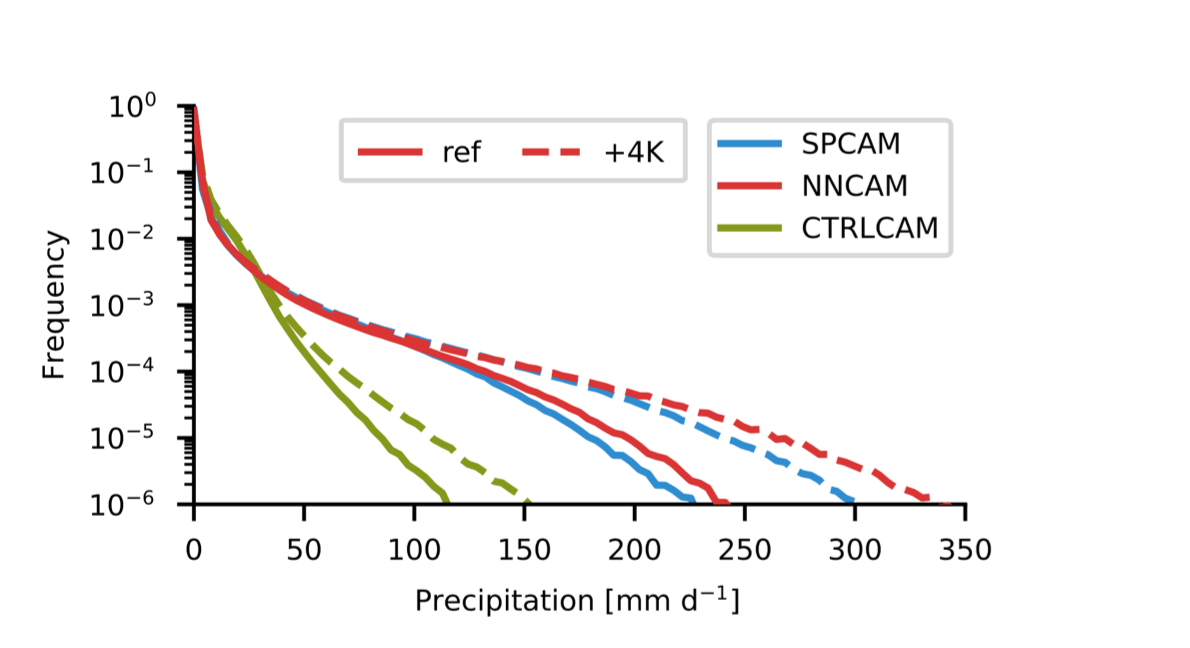}
\caption{Precipitation histogram of time-step (30 min) accumulation. The bin 
width is 3.9 mm$\cdot$d$^{-1}$. Solid lines show the simulations for reference 
sea surface temperature (SST). Dashed lines denote simulations for warming up
by +4-K (see the original Rreference [95]). The neural network in the +4-K case
is NNCAM-ref + 4 K [95].}
\end{center}
\end{figure}

\newpage

\begin{figure}[htp]
\begin{center}
\includegraphics[width=\textwidth]{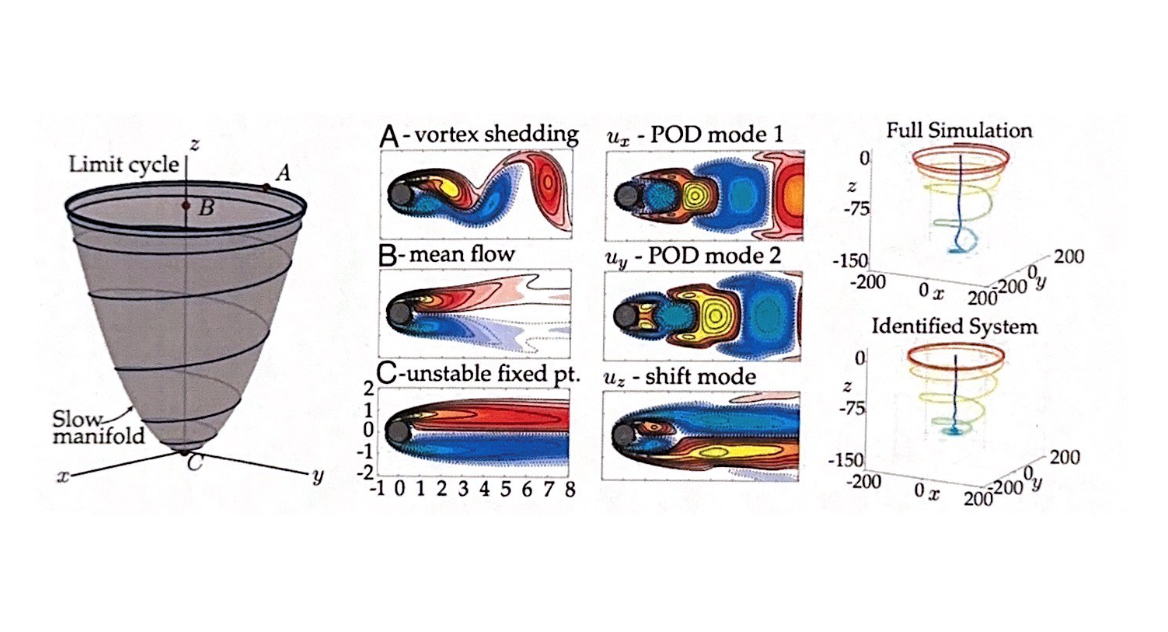}
\caption{The vortex shedding past a cylinder is the result of a Hopf 
bifurcation. Because the Navier-Stokes equations have quadratic nonlinearity, 
one must use a mean-field model with a separation of timescales, where a fast 
mean-field deformation is slave to the slow vortex shedding dynamics. The 
parabolic slow manifold is shown (left), with the unstable fixed point (C), 
mean flow (B), and vortex shedding (A). A proper orthogonal decomposition (POD)
basis and shift mode were used to reduce the dimension of the problem (middle 
right). The identified dynamics closely match the true trajectory in POD 
coordinates, and capture the quadratic nonlinearity and timescales associated 
with the mean-field model [120].}
\end{center}
\end{figure}

\newpage

\begin{figure}[htp]
\begin{center}
\includegraphics[width=\textwidth]{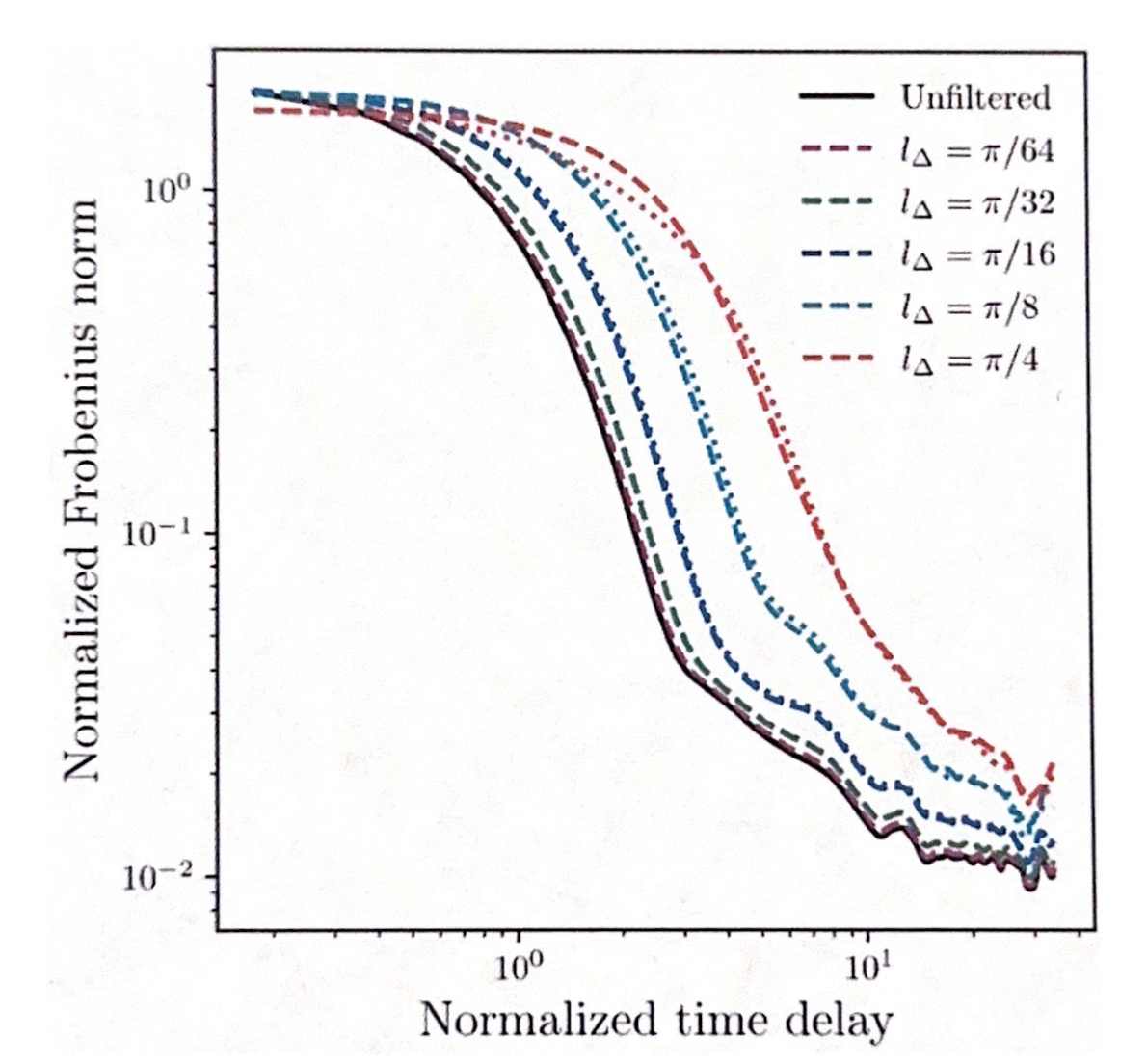}
\caption{Normalized Frobenius norm of the learned memory kernel for observable 
set as a function of normalized time delay. Two types of spatial filters, 
Gaussian and box filters, with various filtering length scales, were applied 
to the physical-space variables [171].}
\end{center}
\end{figure}

\end{document}